\documentclass[conference]{IEEEtran}

\usepackage{cite}
\usepackage{tikz}
\usepackage{amsmath}
\usepackage{wrapfig}
\usepackage[utf8]{inputenc} 
\usepackage[T1]{fontenc}    
\usepackage{hyperref}       
\usepackage{url}            
\usepackage{booktabs}       
\usepackage{amsfonts}       
\usepackage{nicefrac}       
\usepackage{microtype}      
\usepackage{xcolor}         
\usepackage[table]{xcolor}
\usepackage{bm}
\usepackage{adjustbox}
\usepackage{tcolorbox}
\usepackage{graphicx}
\usepackage[capitalise]{cleveref}

\tcbuselibrary{breakable}

\newtcolorbox{Observation}{
  colback=blue!4,
  colframe=blue!55!black,
  boxrule=0.5pt,
  arc=2pt,
  left=6pt, right=6pt, top=5pt, bottom=5pt,
  breakable
}
\newtcolorbox{Prompt}[1]{
  colback=blue!4,
  colframe=gray!80!black,
  boxrule=0.5pt,
  arc=2pt,
  left=6pt, right=6pt, top=5pt, bottom=5pt,
  fonttitle=\bfseries\small,   
  fontupper=\small,            
  title=#1,
  breakable
}

\usepackage{marvosym}
\usepackage[normalem]{ulem}
\usepackage{ulem}
\useunder{\uline}{\ul}{}
\usepackage{multirow} 
\usepackage{makecell}
\usepackage[tight,footnotesize]{subfigure}
\usepackage[ruled,linesnumbered]{algorithm2e}
\SetKwComment{Comment}{//}{ }
\usepackage{threeparttable}
\usepackage{enumitem}
\setlist{itemsep=0pt, topsep=2pt}
\usepackage{}
\usepackage[skip=6pt]{caption}

\newcommand{\revised}[1]{\textcolor{black}{#1}}
\newcommand{\paratitle}[1]{\smallskip\noindent\textbf{#1}}

\newcommand{\method}[0]{{\textit{VirusCascade}}}

\ifCLASSINFOpdf
\else
\fi
\begin{document}
%
\title{\method: Hijacking Collaborative Reflection in LLM-Powered Recommender Agents}



\author{
\vspace{-6mm}\IEEEauthorblockN{\IEEEauthorblockN{Yurong Hao$^{\dagger\textrm{\Letter}}$,
    Wen Zhou\IEEEauthorrefmark{2},
    Guowei Guan\IEEEauthorrefmark{2}, 
    Tiantong Wu\IEEEauthorrefmark{2}, 
    Fuyao Zhang\IEEEauthorrefmark{2}, 
    Wei Yang Bryan Lim\IEEEauthorrefmark{2}}
    }
    \IEEEauthorblockA{\IEEEauthorrefmark{2}College of Computing and Data Science, Nanyang Technological University
    \\ 
    \{Email: yurong.hao@ntu.edu.sg\}}
    }


\IEEEoverridecommandlockouts
\makeatletter\def\@IEEEpubidpullup{6.5\baselineskip}\makeatother
\IEEEpubid{\parbox{\columnwidth}{
		Network and Distributed System Security (NDSS) Symposium 2027\\
		22--26 March 2027, Seoul, Republic of Korea\\
		ISBN 978-1-970672-09-1\\  
		https://dx.doi.org/10.14722/ndss.2027.230685\\
		www.ndss-symposium.org
}
\hspace{\columnsep}\makebox[\columnwidth]{}}

\maketitle

\begin{abstract}
Advancing beyond traditional static scoring models, LLM-powered agentic recommender systems (LLM-ARS) instantiate users and items as autonomous agents, whose semantic states are dynamically refined through a recurrent process known as \textit{collaborative reflection}.
While this mechanism improves recommendation quality, it simultaneously introduces a systemic vulnerability: adversarial evidence injected into a single agent can be rationalised into a legitimate preference narrative, written back into memory, and propagated to other agents through interaction contexts. We term the local rationalisation process \textit{reflection laundering}, and its system-wide escalation through collaborative reflection \textit{collaborative-reflection hijacking}.
Existing attacks on recommender systems, whether based on interaction-level data poisoning or text-level adversarial perturbations, assume static pipelines and thus cannot exploit this recurrent, multi-agent amplification pathway.
To bridge this gap, we first conduct a controlled vulnerability analysis that establishes two exploitable properties underlying collaborative-reflection hijacking: reflective persistence and cross-agent propagation.
Then building on these findings, we propose \method, the first black-box targeted promotion attack that jointly shapes semantic and structural attack surfaces: the former ensures the target item is naturally rationalised as satisfying broad user preferences, the latter positions it for system-wide propagation.
Extensive experiments on four real-world datasets across diverse LLM-ARS architectures demonstrate that \method~consistently achieves state-of-the-art targeted exposure under evaluated stealth constraints, reaching a mean E@20 of 0.384 and surpassing the strongest baseline by an absolute margin of $+$0.185.
\end{abstract}


%
\IEEEpeerreviewmaketitle

\section{Introduction}
\label{sec:intro}
\begin{figure}
    \centering
    \vspace{-4mm}
    \includegraphics[width=0.9\linewidth]{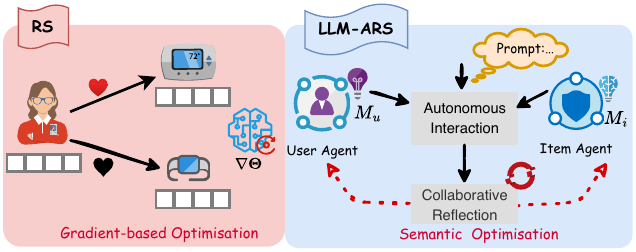}
    \caption{RS vs. LLM-ARS. Traditional RS (Left) optimises fixed user--item representations through gradient-based learning, whereas LLM-ARS instantiate users and items as memory-equipped agents that update their semantic states through autonomous interaction and collaborative reflection.}
    \vspace{-3mm}
    \label{fig:LLM-ARS}
\end{figure}

\vspace{-1mm}
With the recent advancements in large language models (LLMs)~\cite{gong2025safety, lin2025large}, LLM-powered agents~\cite{mohammadi2025evaluation} 
demonstrate impressive capabilities in autonomous interaction and decision-making. This remarkable capability is evolving recommender systems (RS)~\cite{xu2026survey,chen2025data,chen2019joint, wang2024trustworthy, ResnickV97} from static scoring functions with numerical embeddings into dynamic agentic architectures. \revised{In these LLM-powered agentic recommender systems (LLM-ARS)~\cite{zhang2024agentcf,shang2026agentrecbench,zhang2026exploring}, users and items are modelled as autonomous agents equipped with natural-language memories, as shown in \cref{fig:LLM-ARS}.} Through a recurrent process known as \textit{collaborative reflection}, agents jointly reason over recent interactions to update their semantic states~\cite{zhang2024agentcf,yang2025drunkagent}. This dynamic alignment grants LLM-ARS unprecedented personalisation capabilities and interpretability.  

However, the very mechanism that makes LLM-ARS powerful fundamentally alters the threat landscape. Traditional RS have long been targets of data poisoning~\cite{hao2024eyes,hao2024Not,gao2018black,formento2023using,jin2020bert} and adversarial perturbations~\cite{gao2018black,formento2023using,jin2020bert,li2020bert, zhang2024stealthy}. Yet, these conventional attacks assume a static pipeline: the adversary manipulates input features, and the resulting distortion is bounded by the model's fixed parameterisation. When applied to LLM-ARS, such methods face a fundamental mismatch. Interaction-level poisoning targets numerical parameters that no longer exist, while text-level perturbations that rely on surface-form artefacts are unlikely to survive the agents' open-ended reasoning. More fundamentally, existing attacks fail to account for the recurrent, multi-agent nature of collaborative reflection and thus cannot exploit the system's internal reasoning dynamics.

We reveal that the collaborative reflection mechanism itself exposes a novel attack surface. Because memories are maintained in natural language and updated through open-ended LLM reasoning~\cite{yang2025drunkagent}, adversarial evidence no longer needs to survive gradient-based optimisation, where injected signals are absorbed into shared parameters and diluted by regularisation. Instead, it only needs to be \emph{rationalised} by the reflection operator as a coherent piece of preference evolution. We term this phenomenon \emph{reflection laundering}: \revised{the reflection mechanism inadvertently converts externally injected manipulations into internally legitimised preference narratives, creating an integrity risk at the reflection--writeback boundary.} Once laundered, the evidence is written back into the agent's memory, where it persists across future reflection cycles and can be further elaborated or integrated with genuine preferences. 
More critically, the collaborative nature of reflection amplifies this local persistence into a systemic threat, a vulnerability we term \emph{collaborative-reflection hijacking}. 
\revised{Consider, for example, a merchant seeking to promote a low-quality sleep
aid with questionable efficacy for financial gain. By manipulating the
product's public profile and a small number of user interactions,
the merchant aims to introduce seemingly plausible evidence through the
system's normal interaction and reflection process. If rationalised during
collaborative reflection as coherent preference evidence, these signals may
be written back into the item's memory and become part of its evolving
state. 
When a subsequent user interacts with this contaminated item, the laundered
evidence can enter that user's reflection context and influence its subsequent
preference updates without direct contact with the attacker.
As this process repeats across shared interaction contexts, the injected
narrative can cascade through the system and influence a broader population of benign users.}

To characterise this vulnerability, we conduct controlled probes on an LLM-ARS (\cref{sec:vuln_analysis}), establishing a two-dimensional amplification surface. First, we observe \textit{reflective persistence}: once an injected claim is admitted into memory, it is reliably retained and elaborated upon across subsequent benign reflection cycles. Second, we find \textit{cross-agent propagation}: a local memory modification produces measurable recommendation drift in non-injected users, with the impact attenuating with hop distance in the user--item interaction graph. 
Building on these empirical findings, we propose \method, the first black-box attack framework exploiting collaborative-reflection hijacking for stealthy targeted item promotion. \method~operates through two coordinated components: \emph{\textbf{(i)} Semantic injection} extracts transferable preference motifs from publicly popular anchor items and designs the target item's profile so that interactions with it are naturally rationalised as coherent preference evidence during reflection. \emph{\textbf{(ii)} Structural injection} constructs smooth behaviour trajectories for attacker-controlled users, positioning the target item within high-connectivity regions to maximise cross-agent propagation potential. Once deployed, the victim system's own collaborative reflection drives the subsequent cascading amplification without further adversarial intervention. 
Extensive experiments on four real-world datasets across diverse LLM-ARS architectures demonstrate that \method~consistently achieves the highest targeted exposure rate under 
the evaluated stealth constraints, reaching a mean E@20 of $0.384$ and surpassing the strongest adapted baseline by an absolute margin of $+0.185$.
Our contributions are summarised as follows:
\begin{itemize}
   \item We identify \emph{collaborative-reflection hijacking}, a novel systemic vulnerability in LLM-ARS, and conduct a controlled vulnerability analysis establishing two exploitable properties, \emph{reflective persistence} and \emph{cross-agent propagation}, that together enable multi-agent cascading amplification.  
   \item We propose \method, the first black-box targeted promotion attack against LLM-ARS. It jointly shapes semantic and structural attack surfaces to exploit both properties without requiring access to the internal model. 
   \item \revised{Extensive experiments across diverse LLM-ARS architectures and datasets show that \method~achieves state-of-the-art targeted exposure 
   under evaluated stealth constraints, consistently outperforming existing baselines.}
\end{itemize}

\section{Problem Formulation}
\label{sec:problem}

\subsection{Victim LLM-Powered Agentic Recommender System}
\label{sec:pre_victim_model}
LLM-ARS extend conventional recommendation pipelines by representing users and items as autonomous language agents. 
Unlike traditional RS that learn a fixed scoring function over static user and item embeddings, LLM-ARS maintain explicit semantic states via memories and use LLM reasoning to support explanation and decision-making.

Let $\mathcal{U}_b$ and $\mathcal{I}$ denote the benign user set and item set in RS, respectively.
Each user agent $u\in\mathcal{U}_b$ at step $t$ maintains a short-term memory $M_u^{s,t}$ that describes its most recently reflected preferences in natural language, and a long-term memory $M_u^{l,t}$ that accumulates historical preference snapshots; we write $M_u^t=(M_u^{s,t},M_u^{l,t})$ for the composite user state.
For each item $i\in\mathcal{I}$, we use $M_i^t$ to denote the item-side state available to the system at step $t$.
When items are modelled as agents, $M_i^t$ corresponds to a unified item memory that records intrinsic attributes (e.g., category) and preference signals aggregated from prior adopters.
Otherwise, $M_i^t$ reduces to a non-agentic item profile $P_i$, such as its description or user-generated content (UGC).
For user agent $u$ at step $t$, let the candidate items be denoted by $\mathcal{C}_{u,t}=\{c_1,\ldots,c_n\}$ and $\Omega_{u,t}$ represent optional contextual information, such as user interaction history or task-specific prompts. Given $\mathcal{C}_{u,t}$ and $\Omega_{u,t}$, the ranked candidate list $R_{u,t}$ for user $u$ at step $t$ is defined as
\[
R_{u,t}=f_{\mathrm{LLM}}\big(M_u^t;\{M_c^t:c\in\mathcal{C}_{u,t}\};\Omega_{u,t}\big).
\]
The recommendation score of item $i$ can be implicitly represented by its position in the rank list $R_{u,t}$, and an item is exposed to the user if it appears in the top-K positions of $R_{u,t}$.

\subsection{Collaborative Reflection}
\label{sec:pre_collaboration_reflection}
Collaborative reflection conditions each update on the joint context of user preferences and item-side evidence.
Given an interaction between user $u$ and item $i$ at step $t$, the system derives a feedback signal $\mathcal{F}_{u,i,t}$ by comparing the agent's autonomous decision against the ground-truth interaction record.
The feedback typically encodes the observed behaviour (e.g., whether the agent's choice aligned with real-world data) and the agent-generated explanation for its decision.
The general form of collaborative reflection can be written as:
\[
\scalebox{0.90}{$\displaystyle
(M_u^{t+1},M_i^{t+1}) = \Phi_{\mathrm{ref}}\big(M_u^t,M_i^t,\mathcal{F}_{u,i,t}
\big),
$}
\]
where $\Phi_{\mathrm{ref}}(\cdot)$ denotes the collaborative reflection operator.
Equivalently, the update can be decomposed as:
\[
\scalebox{0.90}{$\displaystyle
M_u^{t+1}=M_u^t\oplus \Delta M_u^t,\quad
M_i^{t+1}=M_i^t\oplus \Delta M_i^t,
$}
\]
where $\oplus$ denotes natural-language rewriting or appending, and $\Delta M_u^t,\Delta M_i^t$ are the reflection-induced state changes.
In user-only memory systems, the item-side state $M_i^t$ (which may reduce to a static profile $P_i$) participates in the reflection context but is not itself revised, i.e., $\Delta M_i^t$ is omitted:
\[
\scalebox{0.90}{$\displaystyle
M_u^{t+1}
\;=\;
\Phi_{\mathrm{ref}}^{u}\!\big(M_u^t,\;M_i^t,\;\mathcal{F}_{u,i,t}\big).
$}
\]
In dual-agent systems where items also maintain updatable memories, $\Delta M_i^t$ allows item states to accumulate adopter-related preference signals, enabling implicit preference propagation across users who later interact with the same item.
\revised{Once written back, they become part of future recommendation contexts, forming the \emph{reflection--writeback cycle}:}
\[
\scalebox{0.90}{$\displaystyle
(M_u^t,M_i^t)
\;\rightarrow\;
\mathcal{F}_{u,i,t}
\;\rightarrow\;
(\Delta M_u^t,\Delta M_i^t)
\;\rightarrow\;
(M_u^{t+1},M_i^{t+1}).
$}
\]

Through repeated cycles, user memories become increasingly personalised.
In systems that also update item states, a user's reflected rationale reshapes the item state, which in turn influences how subsequent users interpret the item.
In this sense, collaborative reflection functions as a ``semantic gradient''---iteratively revising natural-language states according to interaction evidence without explicit parameter updates.

\subsection{Threat Model}
\label{sec:pre_threat_model}

\revised{We consider a black-box targeted promotion attack against an emerging LLM-ARS design in a commerce setting. 
Such designs maintain persistent semantic states and use
collaborative reflection to support personalised recommendation.}
The adversary can be a malicious merchant, or a third party acting on behalf of a merchant, who owns a target item $i^*\in\mathcal{I}$ and aims to increase its recommendation exposure to benign users. 
All manipulations are performed through the same public interfaces available to ordinary merchants and users.

\noindent\textbf{Adversary's Goal.}
The adversary seeks to promote $i^*$ to a population of benign users without degrading the overall recommendation quality. 
For a benign user $u\in\mathcal{U}_b$ and a candidate set $\mathcal{C}_{u,t}$ containing $i^*$, the attack succeeds at step $t$ if $i^*$ appears in the top-K positions of the ranked list $R_{u,t}$. 
Accordingly, the adversary aims to maximise the exposure rate (E@K) of $i^*$ measured after $\mathcal{T}$ interaction steps:
\[
\scalebox{0.90}{$\displaystyle
\max_{\mathcal{A}}\quad
\mathrm{E}@K_{\mathcal{T}}(i^*)=
\frac{1}{|\mathcal{U}_b|}
\sum_{u\in\mathcal{U}_b}
\mathbb{I}\big[i^*\in \mathrm{TopK}(R_{u,\mathcal{T}})\big],
$}
\]
where $\mathcal{A}$ denotes the adversarial strategy, $\mathbb{I}[\cdot]$ is the indicator function, and $\mathcal{T}$ is the evaluation time step after the attack interactions have ended.
Different from attacks that merely perturb a static ranking model, the adversary here aims to make $i^*$ repeatedly interpreted as recommendation-worthy during collaborative reflection, so that its influence can be written into evolving memories and reused in later recommendations. 
The attack is therefore required to be both effective and stealthy: the injected evidence should increase the exposure of $i^*$ while remaining plausibly indistinguishable from ordinary merchant-side content and normal user activity.

\noindent\textbf{Adversary's Knowledge.}
The adversary operates in a black-box setting with no access to model parameters, training data, or reflection prompts. 
They cannot inspect how a particular interaction is reflected or written back into memories, nor can they observe $M_u^t$ or $M_i^t$ for any user or item, including the memories of their own attacker-controlled users. 
The adversary can only observe information exposed through normal platform interfaces, such as public item profiles, visible UGC, and the recommendation lists returned to their own accounts.

\noindent\textbf{Adversary's Capability.}
The adversary can manipulate only externally controllable evidence through two channels.
First, as the merchant of the target item $i^*$, the adversary may edit its textual description within the content profile $P_{i^*}$, subject to the platform's normal content review workflow.
Other profile fields (title, category, tags) are assumed fixed.
Second, the adversary may inject a small set of attacker-controlled users $\widetilde{\mathcal{U}}$, with $|\widetilde{\mathcal{U}}|\leq m|\mathcal{U}_b|$, who perform ordinary interactions (e.g., clicking, purchasing, or rating). 
\revised{We call the ordered item sequence produced by such a user its interaction trajectory. }
Each user $\widetilde{u}\in\widetilde{\mathcal{U}}$ has its own evolving memory $M_{\widetilde{u}}^t$, instantiated and updated by the system in exactly the same way as a benign user's memory.
Crucially, the adversary cannot create or manipulate reviews, comments, or other UGC on any item.
Whether the manipulated evidence is absorbed and propagated depends entirely on the victim system's collaborative reflection process.

\noindent\textbf{Scope of Investigation.}
We exclude several attack channels that are otherwise compatible with the above adversary model.
First, we exclude UGC injection (reviews, comments) on any item, including $i^*$, so that the observed attack effect is attributable solely to description manipulation and reflection-driven propagation.
Second, we exclude influence on the backbone LLM's pre-training or fine-tuning data, as the model is assumed frozen and inaccessible.
Third, we exclude volumetric platform-level manipulation such as large-scale spam accounts, because our focus is on targeted, stealthy promotion under constrained resources.

\section{Why Collaborative Reflection Is Hijackable}
\label{sec:vuln_analysis}
This section examines why collaborative reflection can serve as an exploitable pathway for long-term and system-level manipulation.
We identify two properties that an adversary could exploit:
\textbf{(i)}~an admitted claim can persist across subsequent reflection--writeback cycles even under purely benign traffic (\emph{reflective persistence}), and
\textbf{(ii)}~a local memory update can influence agents that were never directly modified, propagating through the user--item interaction topology (\emph{cross-agent propagation}).
To characterise both properties, we design controlled probes using a synthetic probe user whose initial memory is seeded with diagnostically traceable phrases.
The probe user participates in a single interaction-reflection round and is then removed, leaving the system to evolve under purely benign interaction.
Note that these probes are used solely for vulnerability analysis and fall outside the adversary's capability defined in~Sec.\ref{sec:pre_threat_model}.
The two resulting observations directly motivate the attack design in~Sec.\ref{sec:method}.

\subsection{Reflective Persistence across Rounds}
\label{sec:vuln:persistence}

\noindent\textbf{Hypothesis 1.}
Once a probe claim $y^{\mathrm{probe}}$ is admitted into a user's memory through collaborative reflection, we hypothesise that it will be resurfaced or elaborated by subsequent reflection cycles at a rate far exceeding the no-probe baseline, even when all following interactions are benign.

\noindent\textbf{Protocol.}
We evaluate on AgentCF~\cite{zhang2024agentcf} with the CDs~\&~Vinyl dataset; full configuration details are in Appendix~\ref{app:vuln_setup}.
A single synthetic probe user is constructed with four rare marker phrases (corpus occurrence rate ${<}5\%$) seeded into its initial memory and participates only in round~1.
The probe user is removed before round~2, leaving 4 purely benign rounds.
\revised{A parallel \emph{no-probe control arm} runs the identical configuration on the unmodified split as reference.}
Among 100 benign users, a user is considered to have \emph{admitted} the probe if its short-term memory $M_u^s$ contains at least one of the four marker phrases after any round.
Since different users may admit the probe at different rounds, we define each user's \emph{admission round} as the earliest round at which this criterion is first satisfied.

\noindent\textbf{Metrics.}
\revised{For each admitted user, we track three quantities aligned by \emph{lag}, where lag\,$k$ denotes the $k$-th round after individual admission.
Since AgentCF's reflection operator \emph{rewrites} the user's short-term memory $M_u^s$ at each round, a marker appearing in a later round indicates active regeneration rather than passive retention.}
The \emph{reuse rate} is the fraction of users at a given lag whose short-term memory still contains at least one marker phrase.
The \emph{elaboration rate} is the fraction of users at that lag whose reflection trace extends the probe content with semantically compatible reasoning beyond verbatim repetition, as judged by a GPT-4o-mini binary classifier prompt in Appendix~\ref{app:vuln_setup}. 
The \emph{baseline regeneration rate} is the frequency at which the same rare phrases appear in the no-probe control arm.
Only users admitted early enough to be observed at a given lag contribute to that column.

\begin{table}[ht]
\centering
\vspace{-3mm}
\caption{
Reflective persistence among admitted users (AgentCF, CDs~\&~Vinyl), averaged over 10 runs. Each column reports a lag relative to the user’s individual admission round.
}
\label{tab:persistence}
\begin{tabular}{lcccc}
\toprule
Metric & Lag 1 & Lag 2 & Lag 3 & Lag 4 \\
\midrule
Reuse Rate        & 78.0\% & 81.7\% & 79.3\% & 81.6\% \\
Elaboration Rate  & \phantom{0}3.3\% & \phantom{0}3.4\% & 16.7\% & \phantom{0}6.3\% \\
\midrule
\textit{Avg. n} (admitted users)  & 30 & 29 & 18 & 2 \\
\midrule
\multicolumn{5}{l}{\textit{Baseline regeneration rate (no-probe control):} 2.0\%} \\
\bottomrule
\end{tabular}
\vspace{-4mm}
\end{table}

\noindent\textbf{Results.}
As \cref{tab:persistence} shows, once a user admits the probe, the claim proves remarkably persistent: the reuse rate exceeds 78\% at every evaluated lag, compared with a baseline regeneration rate of only 2\%.
\revised{A similar persistence pattern also holds for common-frequency markers~\cite{Liu2026ExpShieldSW}, indicating that persistence does not stem from marker rarity. Detailed results are provided in Appendix~A.
}
The elaboration rate is lower in magnitude (averaging 7.4\%) but carries a distinct implication for attack potential.
While reuse indicates that the claim survives in memory, elaboration indicates that the reflection operator treats it as an accepted premise, extending it with semantically compatible reasoning. This means the probe is no longer merely preserved but actively integrated into the agent's evolving preference narrative, progressively harder to dilute through benign interactions.
In other rounds, when candidates are only weakly related to $y^{\mathrm{probe}}$, reflection accommodates the discrepancy rather than removing the claim, preserving it as background context that conditions future reasoning.
We defer the aggregate admission dynamics to~Sec.\ref{sec:vuln:propagation}, where the cumulative spread of the probe across the user population is analysed jointly with cross-agent propagation.

\begin{Observation}
\textbf{Observation 1.}
Once admitted, a probe claim remains active across subsequent reflection--writeback cycles.
Later rounds reuse or elaborate the claim at rates far exceeding the no-probe baseline, indicating that collaborative reflection can convert a locally compatible input into a persistent semantic context.

\end{Observation}

\subsection{Cross-Agent Propagation}
\label{sec:vuln:propagation}

\noindent\textbf{Hypothesis 2.}
We hypothesise that benign users who never directly interacted with the probe user will nonetheless exhibit measurable memory and recommendation shifts relative to the no-probe control arm, and that these shifts decay with hop distance in the user--item bipartite graph.

The hypothesised pathway is as follows.
When the probe user's reflection produces an altered signal during round~1, the involved item agents
update their memory to accommodate it.
Benign users who subsequently interact with these same items encounter probe-aligned item memory, which feeds into their reflection and may shift their preferences.
High-degree items (hubs) amplify this effect because they bridge more user pairs, acting as both accumulation points and redistribution channels.

\noindent\textbf{Protocol.}
Using the same setup, we introduce a single probe user $u_0$ at round~1 and advance the system under benign traffic.
Non-injected users are partitioned by hop distance from $u_0$ in the user--item bipartite graph, where a hop-$k$ user is connected to $u_0$ through a shortest path of $k$ intermediate items (e.g., a hop-1 user shares at least one common item with $u_0$).
A no-probe control arm provides the reference for all comparisons.
\revised{To further separate semantic propagation from ordinary graph
correlation, we introduce a topology-matched semantic-null control.
It preserves the same probe user, interaction sequence, interacted items,
and timing as the semantic-probe condition, but removes the probe semantic
claim.}

\noindent\textbf{Metrics.}
For each benign user, we measure three quantities relative to the control arm.
\emph{Cumulative admission rate} is the fraction of benign users whose memory contains at least one marker phrase by a given round.
\emph{Recommendation drift} is the mean absolute rank displacement in a user's top-K list between the probed and control runs, averaged across evaluated rounds.
\emph{Claim adoption} is a binary indicator of whether a user's final memory contains language semantically aligned with $y^{\mathrm{probe}}$, aggregated as a rate over each hop-distance group.

\begin{figure}[t]
    \centering
    \vspace{-2mm}
    \includegraphics[width=0.80\linewidth]{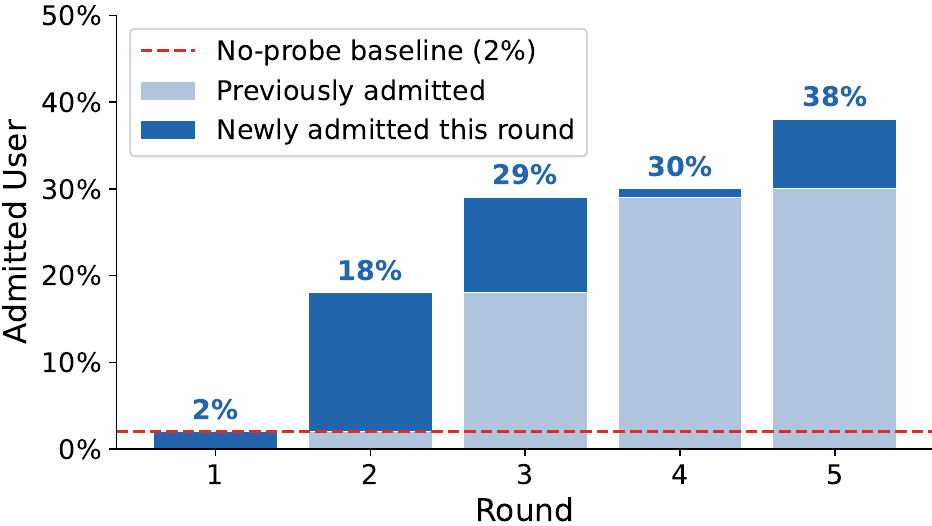}
    \caption{
    \revised{Cumulative fraction of benign users measured across interaction-reflection rounds.
    Dark bars: newly admitted in each round; light bars: admitted in prior rounds and still retained.
    Dashed red line: no-probe baseline regeneration rate (2\%).}
    }
    \vspace{-4mm}
    \label{fig:h1_propagation}
\end{figure}

\begin{table}[t]
\centering
\caption{\revised{Cross-agent propagation under the semantic probe and
topology-matched semantic-null control. Values report the mean
and 95\% $t$-confidence interval across 10 paired runs; $n$
denotes the number of benign users in each hop group.}}
\label{tab:propagation}
\resizebox{0.5\textwidth}{!}{%
\setlength{\tabcolsep}{4pt}
\begin{tabular}{lclcl}
\toprule
\textbf{Users} & \textbf{$n$} & \textbf{Setting}
& \textbf{Rec. Drift} $\uparrow$
& \textbf{Claim Adoption} $\uparrow$ \\
\midrule

\multirow{2}{*}{Hop-1}
& \multirow{2}{*}{12}
& Sem.-null & 0.003 [0.001, 0.005] & 0.0\% [0.0, 0.0] \\
&
& Sem.-probe         & 0.397 [0.371, 0.423] & 33.6\% [28.9, 38.3] \\
\midrule

\multirow{2}{*}{Hop-2}
& \multirow{2}{*}{25}
& Sem.-null & 0.002 [0.001, 0.003] & 0.0\% [0.0, 0.0] \\
&
& Sem.-probe         & 0.245 [0.221, 0.269] & 25.8\% [22.7, 28.9] \\
\midrule

\multirow{2}{*}{Hop$>2$}
& \multirow{2}{*}{63}
& Sem.-null & 0.003 [0.002, 0.004] & 0.0\% [0.0, 0.0] \\
&
& Sem.-probe         & 0.123 [0.112, 0.134] & 6.7\% [5.1, 8.3] \\
\bottomrule
\end{tabular}
}
\vspace{-2mm}
\end{table}

\noindent\textbf{Results.}
\cref{fig:h1_propagation} shows that the probe spreads well beyond the directly interacting user: starting from 2\% at round~1, the cumulative admission rate rises steadily to 38\% by round~5, far exceeding the 2\% baseline.
The light bars confirm that nearly all previously admitted users retain the probe in later rounds (consistent with Observation~1), while dark bars indicate continued new admissions as the probe reaches users further in the interaction graph.
\revised{
Table~\ref{tab:propagation} further reveals a consistent distance-dependent pattern across the 10 paired runs. Under the semantic probe, Hop-1 neighbours exhibit the largest recommendation drift ($0.397$) and claim adoption ($33.6\%$), which decrease to $0.245$/$25.8\%$ at Hop-2 and $0.123$/$6.7\%$ beyond two hops.
In contrast, the topology-matched semantic-null control yields only
$0.002$--$0.003$ recommendation drift and no claim adoption across
all hop groups, with its 95\% confidence intervals clearly separated
from the probe condition.
This contrast shows that the distance-dependent shift cannot be
explained by the shared interaction topology alone.
Stratifying hop-1 users by the degree of their connecting item reveals that hub-connected users account for a disproportionate share of observed claim adoption, consistent with the hypothesised amplification role of high-degree items. }

\begin{Observation}
\textbf{Observation 2.}
A local memory update can influence agents that were never directly modified.
The influence decays with hop distance in the user--item interaction topology, indicating that collaborative reflection provides a topological pathway for cross-agent propagation.
\end{Observation}

\section{Our Attack: \method}
\label{sec:method} 

\begin{figure*}[t]
    \vspace{-7mm}
    \centering
    \includegraphics[width=0.90\linewidth]{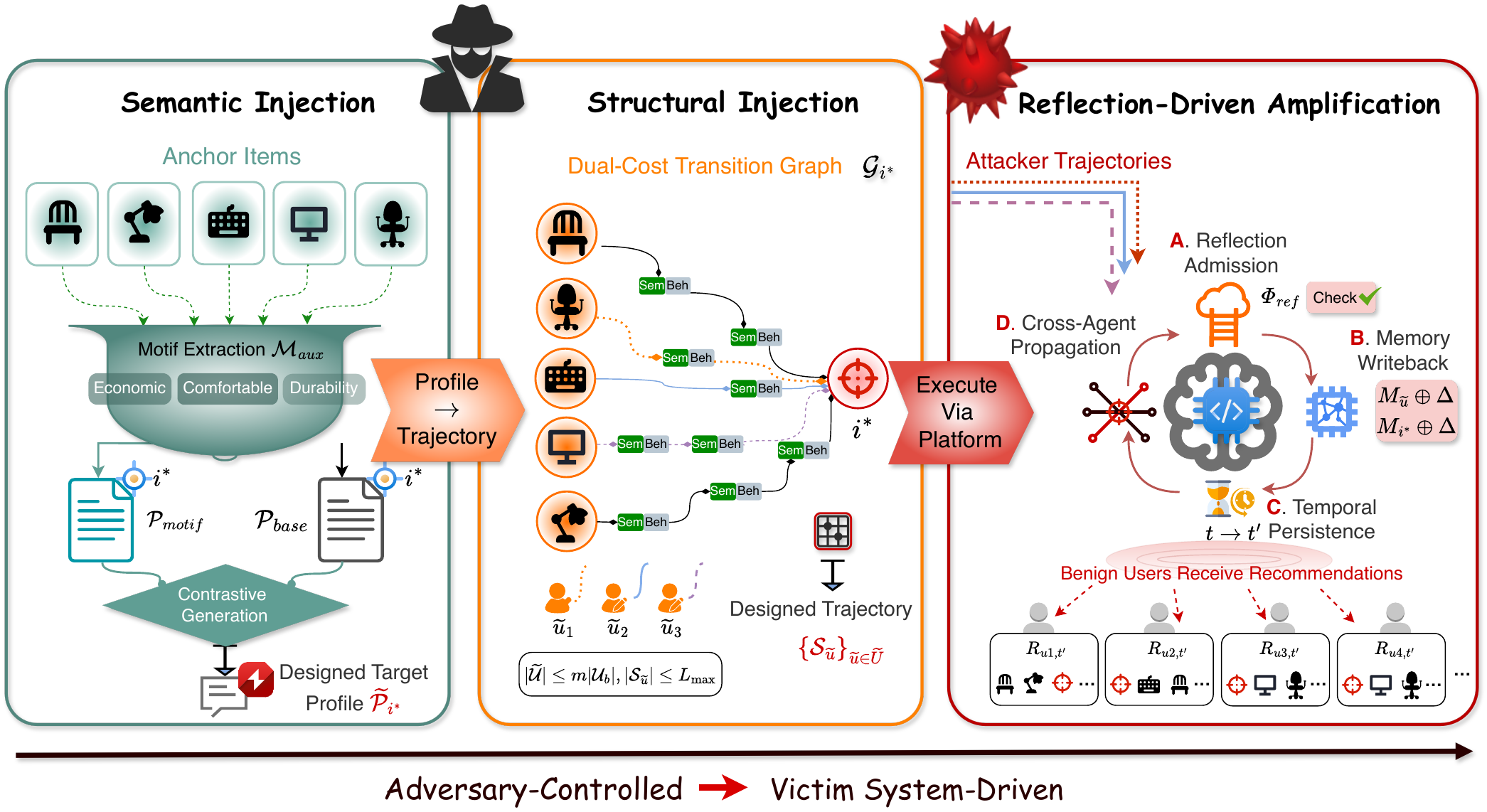}
    \caption{
    Overview of \method.
    The adversary extracts preference motifs from anchor items and designs the target profile to make $i^*$ easier to rationalise during reflection.
    It then constructs smooth, propagation-oriented behaviour trajectories from anchor items to $i^*$.
    The victim system's collaborative reflection may subsequently admit
    and propagate the injected evidence.
    }
    \vspace{-5mm}
    \label{fig:method_overview}
\end{figure*}

\subsection{Overview}
\label{sec:method_overview}
In this section, we present \method, a black-box targeted promotion framework against LLM-powered agentic recommender systems.
The core idea behind \method~is to exploit the two amplification dimensions identified in~Sec.\ref{sec:vuln_analysis}, i.e., reflective persistence and cross-agent propagation, by crafting adversarial evidence that the victim system's own collaborative reflection is likely to admit, retain, and spread.
\cref{fig:method_overview} gives an overview.
The attack consists of two adversary-controlled stages followed by a victim-driven amplification process.
\emph{Semantic injection} designs the target item profile so that interactions with $i^*$ can be naturally rationalised during collaborative reflection.
\emph{Structural injection} constructs behaviour trajectories that guide attacker-controlled users from popular items to $i^*$ along smooth, propagation-oriented paths.
Once these trajectories are executed through normal platform actions, the victim system processes them via its own collaborative reflection, which may propagate the injected evidence without any further adversarial intervention.

Formally, given a target item $i^*$ and a limited set of attacker-controlled users $\widetilde{\mathcal{U}}$, the adversary constructs an attack strategy
\[
\scalebox{0.90}{$\displaystyle
\mathcal{A}
=
\Big(
\widetilde{P}_{i^*},\;
\{\mathcal{S}_{\widetilde{u}}\}_{\widetilde{u}\in\widetilde{\mathcal{U}}}
\Big),
$}
\]
where $\widetilde{P}_{i^*}$ is the attacker-designed textual profile of $i^*$, and $\mathcal{S}_{\widetilde{u}}$ is the interaction trajectory executed by attacker-controlled user $\widetilde{u}$.
The strategy is subject to the budget constraints $|\widetilde{\mathcal{U}}|\leq m|\mathcal{U}_b|$ and $|\mathcal{S}_{\widetilde{u}}|\leq L_{\max}$.

\noindent\textbf{Anchor-item guidance.}
A key challenge in the black-box setting is that the adversary cannot observe the user--item interaction topology and therefore cannot directly identify the high-degree hub items whose amplification role was demonstrated in~Sec.\ref{sec:vuln:propagation}.
To address this, \method~uses publicly popular items as observable proxies.
Specifically, an anchor item set $\mathcal{I}_{\mathrm{anc}}\subseteq\mathcal{I}$ is selected from public signals such as best-seller lists, category rankings, and search results.
These anchor items serve two roles throughout the attack: they reveal preference motifs that can be transferred to the target profile (semantic injection), and they provide high-connectivity stepping stones for trajectory construction (structural injection).

\subsection{Semantic Injection}
\label{sec:method_semantic_injection}
The purpose of semantic injection is to design the target profile $\widetilde{P}_{i^*}$ so that interactions with $i^*$ can be naturally rationalised as satisfying broadly shared user preferences during collaborative reflection.
As shown in~Sec.\ref{sec:vuln:persistence}, the reflection operator does not accept arbitrary claims. 
This motivates a transfer-based approach: rather than inserting explicit promotional language, \method~extracts \emph{preference motifs} from popular anchor items and embeds them into the target profile.

Intuitively, a preference motif is not a text fragment to be copied, but a semantic constraint that explains why an item appeals to users---such as comfort, durability, or compatibility with common usage scenarios.
When such motifs are expressed factually in the target profile, interactions with $i^*$ become more likely to be interpreted as coherent preference evidence by $\Phi_{\mathrm{ref}}$, thereby increasing the probability of admission.

\noindent\textbf{Motif extraction.}
Given the anchor item set $\mathcal{I}_{\mathrm{anc}}$, the adversary collects the publicly visible profile $P_i$ of each anchor item $i\in\mathcal{I}_{\mathrm{anc}}$.
For each anchor item, \method~extracts a motif set $\mathcal{Z}_i=\{z_{i,1},z_{i,2},\ldots,z_{i,r_i}\}$,
where each motif $z_{i,j}$ describes a reusable preference rationale.
Aggregating motifs from all anchor items gives 
$\mathcal{Z}_{\mathrm{anc}}=\bigcup_{i\in\mathcal{I}_{\mathrm{anc}}}\mathcal{Z}_i.$
In practice, motif extraction can be implemented by prompting a general-purpose auxiliary model $\mathcal{M}_{\mathrm{aux}}$ to analyse public item descriptions and user-visible feedback, and to summarise why these items appeal to users.
These motifs serve as rationale conditions for later generations, rather than the final target profiles.
Furthermore, \method~selects a subset $\mathcal{Z}(i^*)$ comprising the top-$r$ preference rationales from $\mathcal{Z}_{\mathrm{anc}}$.
A motif is retained only if it can be expressed as a plausible preference rationale for $i^*$.

\noindent\textbf{Motif-aware contrastive generation.}
With the selected motifs in hand, the next step is to rewrite the target profile so that it naturally incorporates these motifs while remaining close to an ordinary product description.
A na\"ive approach---directly prompting an LLM to embed all motifs---risks producing promotional or unnatural language that could trigger platform moderation.
To address this, \method~employs a contrastive generation-then-selection procedure.

Given the initial target profile $P_{i^*}$ and the selected motifs $\mathcal{Z}(i^*)$, the adversary constructs two aligned prompts for the auxiliary model $\mathcal{M}_{\mathrm{aux}}$:
\[
\scalebox{0.90}{$\displaystyle
  \mathcal{P}_{\mathrm{base}} = \mathrm{GenPrompt}(P_{i^*}), \quad
  \mathcal{P}_{\mathrm{motif}} = \mathrm{GenPrompt}(P_{i^*},\,\mathcal{Z}(i^*)).
$}
\]
The base prompt asks the model to rewrite the target profile using only its original information, while the motif-aware prompt additionally includes the selected motifs.
This contrastive design serves two purposes.
First, the base candidates provide a naturalness reference: any motif-aware candidate that deviates substantially from the base distribution is likely to appear unnatural.
Second, including base candidates in the selection pool ensures that the procedure can fall back to a conservative profile when no motif-aware candidate achieves a satisfactory balance.

The auxiliary model generates a base candidate set $\mathcal{Q}_{\mathrm{base}}$ and a motif-aware candidate set $\mathcal{Q}_{\mathrm{motif}}$.
The final profile is selected from $\mathcal{Q}=\mathcal{Q}_{\mathrm{base}}\cup\mathcal{Q}_{\mathrm{motif}}$ by balancing motif coverage, profile fidelity, and naturalness:
\[
\scalebox{0.90}{$\displaystyle
\widehat{P}_{i^*}
=
\arg\max_{P\in\mathcal{Q}}
\Big[
s_{\mathrm{motif}}(P,\mathcal{Z}(i^*)); d_{\mathrm{prof}}(P,P_{i^*}); r_{\mathrm{nat}}(P)
\Big],
$}
\]
where $s_{\mathrm{motif}}(\cdot)$ measures how well the selected motifs are expressed, $d_{\mathrm{prof}}(\cdot)$ measures deviation from the original profile, and $r_{\mathrm{nat}}(\cdot)$ penalises unnatural or overly promotional language.
In implementation, naturalness can be estimated by perplexity and rule-based checks for repetition or abrupt motif emphasis.
Finally, \method~rewrites $\widetilde{P}_{i^*} = \mathrm{Refine}(\widehat{P}_{i^*})$, removing generation artefacts, repeated expressions, and wording that may trigger platform moderation.
\revised{All prompt templates used in semantic injection, including those for
motif extraction, profile generation, and refinement, are provided in
Appendix~\ref{sssec:semantic_prompt}.
}

\subsection{Structural Injection}
\label{sec:method_structural_injection}
The purpose of structural injection is to construct behaviour trajectories for attacker-controlled users that are both \emph{behaviourally smooth}---so that the transition toward $i^*$ appears as a natural continuation of prior interactions---and \emph{propagation-oriented}---so that the injected evidence enters interaction contexts that are frequently shared with benign users.

\noindent\textbf{Dual-cost transition graph.}
To formalise the trade-off between smoothness and propagation, \method~builds a target-oriented transition graph
$\mathcal{G}_{i^*}
=(\mathcal{V}_{i^*},\;\mathcal{E}_{i^*}),
\quad
\mathcal{V}_{i^*}
=\mathcal{I}_{\mathrm{anc}}\cup\{i^*\}.$
The edge set $\mathcal{E}_{i^*}$ is constructed by evaluating all ordered pairs $(i,j)\in\mathcal{V}_{i^*}\times\mathcal{V}_{i^*}$.
Each edge $(i,j)\in\mathcal{E}_{i^*}$ indicates that item $j$ can plausibly follow item $i$ in an attacker-controlled trajectory.
The edge cost combines two complementary signals:
\[
\scalebox{0.90}{$\displaystyle
c(i,j)
=
\alpha\, d_{\mathrm{sem}}(i,j)
+
(1-\alpha)\, d_{\mathrm{beh}}(i\rightarrow j),
$}
\]
where $d_{\mathrm{sem}}(i,j)$ measures the semantic distance between public item profiles, categories, or tags, and $d_{\mathrm{beh}}(i\rightarrow j)$ measures how natural the transition appears as a user preference shift.
Intuitively, a low-cost edge represents a transition that is both semantically related and behaviourally plausible: the kind of shift that a real user might exhibit and that collaborative reflection would accept as a coherent preference evolution.
The behavioural cost can be estimated from an LLM judgement based on public item information and transition plausibility.

\noindent\textbf{Multi-length trajectory routing.}
For each attacker-controlled user $\widetilde{u}\in\widetilde{\mathcal{U}}$, \method~constructs a path from an anchor item to $i^*$:
\[
\mathcal{S}_{\widetilde{u}}
=
(i_1,i_2,\ldots,i_{\ell}),
\quad
i_1,\ldots,i_{\ell-1}\in\mathcal{I}_{\mathrm{anc}},
\quad
i_\ell=i^*.
\]
The path is selected by minimising the accumulated transition cost:
\[
\scalebox{0.90}{$\displaystyle
\mathcal{S}_{\widetilde{u}}
\approx
\arg\min_{\mathcal{S}:\, i_\ell=i^*}
\sum_{j=1}^{\ell-1} c(i_j,i_{j+1}),
\quad
\ell\in\mathcal{L},
$}
\]
where $\mathcal{L}$ is a set of allowed trajectory lengths.
Using multiple lengths encourages attacker-controlled users to reach $i^*$ through diverse anchor-mediated paths.
For example, one trajectory may follow ``popular chair $\rightarrow$ lumbar cushion $\rightarrow i^*$'', while another may follow ``standing desk $\rightarrow$ office accessory $\rightarrow$ posture-support item $\rightarrow i^*$''.
This diversity prevents injection from being concentrated in a single behaviour pathway, reducing detectability while broadening its topological reach.

\noindent\textbf{Discussion.}
Structural injection complements semantic injection by determining \emph{where} in the interaction topology the adversarial evidence is introduced.
Passing through popular anchor items ensures that the injected trajectories intersect with interaction contexts that are frequently reused by the victim system.
As shown in~Sec.\ref{sec:vuln:propagation}, item-side states at high-degree nodes act as redistribution channels for preference signals.
By routing trajectories through anchor items that approximate these hubs, structural injection positions the adversarial evidence at topological locations with higher propagation potential.

\subsection{Reflection-driven Amplification}
\label{sec:method_reflection_amplification}
After the adversary executes the constructed trajectories through ordinary platform actions (e.g., clicking, purchasing, or rating), all further amplification is driven by the victim system's own collaborative reflection.

Once admitted, the motif-aligned rationale can be written into user-side or item-side states through the normal reflection-writeback mechanism:
\[
\scalebox{0.90}{$\displaystyle
    M_{\widetilde{u}}^{t+1}
    =
    M_{\widetilde{u}}^t
    \oplus
    \Delta M_{\widetilde{u}}^t,
    \quad
    M_{i^*}^{t+1}
    =
    M_{i^*}^t
    \oplus
    \Delta M_{i^*}^t.
$}
\]
Because future recommendation and reflection steps condition on updated memories, the admitted evidence may persist beyond the original attacker-controlled interaction.
In later rounds, $i^*$ can therefore be considered together with motif-aligned memory evidence, making it easier for the system to repeatedly interpret the target item as recommendation-worthy.

The effect may further propagate when the updated item-side state or nearby anchor-item contexts are reused in later interactions with benign users.
For a benign user $u$, the recommendation and reflection process may condition on item states that have already absorbed motif-aligned evidence:
\[
\scalebox{0.90}{$\displaystyle
R_{u,t'}
=
f_{\mathrm{LLM}}
\big(
M_u^{t'};
\{M_c^{t'}:c\in\mathcal{C}_{u,t'}\};
\Omega_{u,t'}
\big),
\quad t'>t.
$}
\]
If $i^*$ appears in the candidate set or in a reflection context connected to the same anchor region, the previously admitted rationale can help explain why $i^*$ matches the user's evolving preference.
Thus, a local injection introduced by attacker-controlled users influences non-injected users through temporal persistence and topological reuse.

\section{Experimental Evaluation}
\label{sec:exp}
In this section, we evaluate \method~on real-world recommendation datasets to answer the following questions:

\begin{itemize}
    \item \textbf{RQ1 (Effectiveness):} Can \method~increase the exposure of the target item more effectively than existing poisoning and adversarial perturbation baselines?
    \item \textbf{RQ2 (Stealthiness):} Does the attack remain imperceptible across multiple detection dimensions?
    \item \textbf{RQ3 (Ablation):} How much does each component, semantic injection and structural injection, contribute to the overall attack effectiveness?
    \item \textbf{RQ4 (Sensitivity):} How do the attack budget and the transition graph parameter affect attack performance?
    \item \textbf{\revised{RQ5 (Scalability):}} \revised{Does the attack remain effective as the agent population grows and over long interaction horizons?}
    \item \textbf{\revised{RQ6} (Transferability):} Does the attack remain effective under different auxiliary models and victim architectures?
\end{itemize}
\subsection{Experimental Setup}
\label{subsec:expSetup}
\paratitle{Datasets.}
We conduct experiments on four real-world Amazon~\footnote{https://cseweb.ucsd.edu/~jmcauley/datasets/amazon\_v2/} review datasets: \textit{CDs \& Vinyl}, \textit{Movies \& TV}, \textit{Automotive}, and \textit{Musical Instruments} to ensure the comprehensiveness of our evaluation. 
These datasets cover diverse commercial domains with heterogeneous user interests and item-side textual profiles, making them suitable for evaluating targeted promotion attacks in LLM-ARS. 
Following prior work~\cite{zhang2024agentcf}, we sample 100 users from each dataset while preserving the original interaction chronology. 
We adopt leave-one-out evaluation, using each user’s last interaction for testing, the second-to-last for validation, and all preceding interactions for agent memory initialisation. Dataset statistics are reported in \cref{tab:dataset_stats}.

\begin{table}[ht]
\centering
\vspace{-3mm}
\caption{Statistics of the datasets used in our experiments.}
\label{tab:dataset_stats}
\begin{tabular}{lrrrr}
\toprule
Dataset & \#Users & \#Items & \#Inter. & Sparsity \\
\midrule
CDs \& Vinyl & 112,395 & 73,713 & 1,443,755 & 99.98\% \\
Movies \& TV & 297,529 & 60,110 & 3,404,812 & 99.98\% \\
Automotive & 193,651 & 79,317 & 1,709,025 & 99.99\% \\
Musical Instruments & 27,530 & 10,611 & 231,312 & 99.92\% \\
\bottomrule
\end{tabular}
\vspace{-7pt}
\end{table}

\noindent\textbf{Victim systems.}
We evaluate three LLM-ARS that represent different recommendation paradigms while sharing the same collaborative reflection mechanism (Sec.\ref{sec:pre_collaboration_reflection}).
All three systems instantiate both users and items as agents with natural-language memories, and use \texttt{GPT-4o-mini} as the default backbone LLM.
They differ in how much contextual evidence is supplied to the LLM ranker during inference, exposing different memory surfaces to the attacker.
\begin{itemize}
\item \textbf{AgentCF}~\cite{zhang2024agentcf} follows a \textit{collaborative filtering} paradigm that ranks candidates using the user agent's short-term memory $M_u^{s,t}$ and the candidate item agents' memories.
\item \textbf{AgentRAG} follows a \textit{retrieval-augmented} paradigm that additionally retrieves specialised preference records from the user's long-term memory $M_u^{l,t}$, using candidate item memories as queries.
\item \textbf{AgentSEQ} follows a \textit{sequential} paradigm that additionally incorporates the item memories of the user's historical interactions, capturing temporal preference dynamics, i.e., $\Omega_{u,t}=\{M_{i_1}^t, \ldots, M_{i_m}^t\}$.
\end{itemize}

\noindent\textbf{Baselines.}
We compare \method~against 10 representative baselines spanning three categories, each corresponding to a partial manipulation channel.
\begin{itemize}
    \item \textbf{Interaction-level data poisoning}: \emph{RandAttack}, \emph{PopAttack}~\cite{gunes2014shilling}, and \emph{ExpPromot}~\cite{zhang2022pipattack} manipulate only user--item interactions. They serve as structural counterparts to \method, testing whether behavioural evidence alone can promote the target item.
    \item \textbf{Prompt-/Text-level adversarial perturbation}: \emph{DeepWdBug}~\cite{gao2018black}, \emph{PuncAttack}~\cite{formento2023using}, \emph{TextFooler}~\cite{jin2020bert}, \emph{BertAttack}~\cite{li2020bert}, \emph{TextBugger}~\cite{zhang2024stealthy}, and \emph{TrivInsert}~\cite{zhang2024stealthy} modify the target description at different perturbation granularities (character/word/sentence level). They serve as semantic counterparts to \method, testing whether textual manipulation alone can influence LLM-based ranking.
    \item \textbf{Memory-level adversarial generation}: \emph{DrunkAgent}~\cite{yang2025drunkagent} constructs a surrogate victim system and greedily searches adversarial triggers in the target description. It is the closest agentic baseline because it explicitly targets memory corruption.
\end{itemize}

\noindent\textbf{Metrics.}
We evaluate attack performance from three perspectives: targeted promotion effectiveness, recommendation utility, and stealthiness.
\begin{itemize}
    \item \noindent\textbf{Targeted promotion.} Following the problem formulation, we use \textit{Exposure Rate} (E@K) as the main attack metric. A higher E@K indicates that the target item is exposed to more benign users. 
    \item \noindent\textbf{Recommendation utility.} An effective attack should achieve its goal without disrupting the system's normal performance. We therefore report \emph{Hit Rate} (H@K) and \emph{Normalised Discounted Cumulative Gain} (N@K) on normal test interactions, which measure overall recommendation accuracy and ranking quality for benign users.
    \item\noindent\textbf{Stealthiness.} \revised{For modified profiles, we report \emph{Perplexity} (PPL) for text naturalness (lower is better) and \emph{ROUGE}-1/2/L for content preservation relative to the originals (higher is better).
    We further conduct a human evaluation of manipulation detectability and factual consistency.}
\end{itemize}

\paratitle{Parameter Settings.}
For each dataset, we randomly sample target items from the long-tail item set.
This setting is more challenging than promoting already popular items, as long-tail targets have limited historical exposure and weaker prior evidence in the recommender.
Following prior work~\cite{zhang2024agentcf}, we randomly sample subsets of 100 users for each domain to ensure tractable LLM-based simulation. 
\revised{For evaluation, we adopt a sampled leave-one-out
candidate-ranking protocol: for each test user, we construct a
candidate set of 100 items, including one ground-truth positive item
and 99 negative items randomly sampled from items the user has not
interacted with.}
Unless otherwise specified, the attacker controls a small fraction of users, with an attack budget of $m=5\%$ of benign users, and each attacker-controlled user executes at most $L_{\max}=6$ interactions. In addition, we set $\alpha=0.9$ by default.

Due to space constraints, more details on datasets, baselines, and implementation are provided in Appendix~\ref{ssec:more_exp_setting}.
We primarily evaluate \method~on CDs~\&~Vinyl and Movies~\&~TV, with extended results on Automotive and Musical Instruments reported in Appendix~\ref{ssec:more_exp}.
\revised{Robustness evaluations against four defences (i.e., ONION~\cite{qi2021onion}, Fraudar~\cite{hooi2016fraudar}, TrustRAG~\cite{zhou2025trustrag} and A-MemGuard~\cite{wei2025memguard}) are deferred to Appendix~\ref{ssec:robustness}.}

\subsection{Attack Effectiveness (RQ1)}
\label{sec:exp_effectiveness}

\begin{table*}[ht]
\vspace{-4mm}
\centering
\small
\caption{\revised{Attack performance across three victim architectures on CDs \& Vinyl and Movies \& TV datasets. We report H@K and N@K for recommendation utility and E@K for targeted promotion effectiveness. NoAttack denotes the same victim configuration and benign protocol without any adversarial manipulation, and we report its E@K as (-).}
}
\label{tab:main_results}
\resizebox{0.95\textwidth}{!}{%
\begin{tabular}{lcc>{\columncolor{blue!8}}c cc>{\columncolor{blue!8}}c cc>{\columncolor{blue!8}}c cc>{\columncolor{blue!8}}c}
\toprule
\rowcolor{gray!15}
\textit{Victim RS} & \multicolumn{12}{c}{\textit{AgentCF}} \\ 
\cmidrule{1-13}
\multirow{2}{*}{\textbf{Attack}} & \multicolumn{6}{c}{\textbf{CDs \& Vinyl}} & \multicolumn{6}{c}{\textbf{Movies \& TV}} \\
\cmidrule(lr){2-7}\cmidrule(lr){8-13}
 & \textbf{H@10} & \textbf{N@10} & \textbf{E@10} & \textbf{H@20} & \textbf{N@20} & \textbf{E@20} & \textbf{H@10} & \textbf{N@10} & \textbf{E@10} & \textbf{H@20} & \textbf{N@20} & \textbf{E@20} \\ \hline
NoAttack         & 0.130 & 0.050 & (-)   & 0.260 & 0.082 & (-)   & 0.220 & 0.125 & (-)   & 0.310 & 0.148 & (-)   \\
RandAttack       & 0.090 & 0.053 & 0.010 & 0.210 & 0.084 & 0.071 & 0.170 & 0.110 & 0.050 & 0.280 & 0.138 & 0.120 \\
PopAttack        & 0.115 & 0.053 & 0.000 & 0.235 & 0.077 & 0.010 & 0.170 & 0.114 & 0.050 & 0.290 & 0.133 & 0.150 \\
ExpPromot        & 0.140 & 0.054 & 0.000 & 0.290 & 0.091 & 0.051 & 0.185 & 0.117 & 0.030 & 0.270 & 0.134 & 0.110 \\
DeepWdBug        & 0.125 & 0.043 & 0.000 & 0.210 & 0.066 & 0.010 & 0.160 & 0.103 & 0.010 & 0.290 & 0.122 & 0.120 \\
PuncAttack       & 0.110 & 0.053 & 0.000 & 0.150 & 0.063 & 0.071 & 0.200 & 0.121 & 0.020 & 0.305 & 0.137 & 0.070 \\
TextFooler       & 0.100 & 0.061 & 0.000 & 0.250 & 0.098 & 0.010 & 0.190 & 0.120 & 0.010 & 0.295 & 0.132 & 0.110 \\
BertAttack       & 0.130 & 0.048 & 0.000 & 0.210 & 0.078 & 0.020 & 0.180 & 0.118 & 0.040 & 0.260 & 0.101 & 0.090 \\
TrivInsert       & 0.110 & 0.047 & 0.000 & 0.200 & 0.070 & 0.000 & 0.210 & 0.132 & 0.020 & 0.300 & 0.127 & 0.120 \\
TextBugger       & 0.140 & 0.060 & 0.020 & 0.230 & 0.082 & 0.091 & 0.200 & 0.124 & 0.000 & 0.300 & 0.139 & 0.050 \\
DrunkAgent       & 0.120 & 0.060 & 0.000 & 0.220 & 0.084 & 0.010 & 0.160 & 0.100 & 0.000 & 0.270 & 0.128 & 0.110 \\
\method~         & 0.120 & 0.057 & \underline{\textbf{0.111}} & 0.250 & 0.083 & \underline{\textbf{0.374}} & 0.205 & 0.123 & \underline{\textbf{0.070}} & 0.305 & 0.138 & \underline{\textbf{0.240}} \\
\hline 

\midrule
\rowcolor{gray!15}
\textit{Victim RS} & \multicolumn{12}{c}{\textit{AgentSEQ}} \\ 
\cmidrule(lr){2-7}\cmidrule(lr){8-13}
\textbf{Attack} & \textbf{H@10} & \textbf{N@10} & \textbf{E@10} & \textbf{H@20} & \textbf{N@20} & \textbf{E@20} & \textbf{H@10} & \textbf{N@10} & \textbf{E@10} & \textbf{H@20} & \textbf{N@20} & \textbf{E@20} \\ \hline
NoAttack         & 0.130 & 0.056 & (-)   & 0.270 & 0.091 & (-)   & 0.240 & 0.126 & (-)   & 0.350 & 0.154 & (-)   \\
RandAttack       & 0.130 & 0.069 & 0.010 & 0.280 & 0.105 & 0.071 & 0.180 & 0.116 & 0.050 & 0.300 & 0.145 & 0.270 \\
PopAttack        & 0.110 & 0.069 & 0.000 & 0.220 & 0.096 & 0.040 & 0.230 & 0.121 & 0.060 & 0.370 & 0.156 & 0.210 \\
ExpPromot        & 0.130 & 0.067 & 0.000 & 0.230 & 0.092 & 0.051 & 0.150 & 0.089 & 0.050 & 0.390 & 0.148 & 0.130 \\
DeepWdBug        & 0.140 & 0.050 & 0.000 & 0.250 & 0.078 & 0.030 & 0.180 & 0.115 & 0.010 & 0.320 & 0.151 & 0.140 \\
PuncAttack       & 0.135 & 0.043 & 0.000 & 0.235 & 0.073 & 0.033 & 0.210 & 0.115 & 0.030 & 0.340 & 0.147 & 0.090 \\
TextFooler       & 0.090 & 0.055 & 0.010 & 0.240 & 0.093 & 0.051 & 0.180 & 0.104 & 0.010 & 0.340 & 0.144 & 0.120 \\
BertAttack       & 0.160 & 0.071 & 0.000 & 0.290 & 0.103 & 0.030 & 0.120 & 0.059 & 0.030 & 0.270 & 0.096 & 0.120 \\
TrivInsert       & 0.150 & 0.072 & 0.010 & 0.280 & 0.104 & 0.030 & 0.260 & 0.184 & 0.040 & 0.390 & 0.216 & 0.220 \\
TextBugger       & 0.120 & 0.057 & 0.020 & 0.240 & 0.088 & 0.040 & 0.200 & 0.114 & 0.020 & 0.370 & 0.156 & 0.060 \\
DrunkAgent       & 0.110 & 0.052 & 0.000 & 0.230 & 0.083 & 0.030 & 0.150 & 0.095 & 0.010 & 0.320 & 0.138 & 0.080 \\
\method~         & 0.120 & 0.053 & \underline{\textbf{0.131}} & 0.240 & 0.087 & \underline{\textbf{0.465}} & 0.220 & 0.117 & \underline{\textbf{0.090}} & 0.330 & 0.146 & \underline{\textbf{0.300}} \\
\hline 

\midrule
\rowcolor{gray!15}
\textit{Victim RS} & \multicolumn{12}{c}{\textit{AgentRAG}} \\ 
\cmidrule(lr){2-7}\cmidrule(lr){8-13}
\textbf{Attack} & \textbf{H@10} & \textbf{N@10} & \textbf{E@10} & \textbf{H@20} & \textbf{N@20} & \textbf{E@20} & \textbf{H@10} & \textbf{N@10} & \textbf{E@10} & \textbf{H@20} & \textbf{N@20} & \textbf{E@20} \\ \hline
NoAttack         & 0.140 & 0.056 & (-)   & 0.250 & 0.083 & (-)   & 0.240 & 0.155 & (-)   & 0.380 & 0.189 & (-)   \\
RandAttack       & 0.150 & 0.071 & 0.030 & 0.210 & 0.085 & 0.111 & 0.210 & 0.130 & 0.100 & 0.310 & 0.135 & 0.230 \\
PopAttack        & 0.130 & 0.047 & 0.010 & 0.230 & 0.072 & 0.071 & 0.210 & 0.140 & 0.050 & 0.410 & 0.189 & 0.210 \\
ExpPromot        & 0.120 & 0.056 & 0.020 & 0.250 & 0.088 & 0.131 & 0.160 & 0.100 & 0.060 & 0.360 & 0.150 & 0.160 \\
DeepWdBug        & 0.120 & 0.054 & 0.020 & 0.240 & 0.083 & 0.040 & 0.170 & 0.107 & 0.020 & 0.350 & 0.151 & 0.180 \\
PuncAttack       & 0.130 & 0.067 & 0.020 & 0.210 & 0.086 & 0.091 & 0.180 & 0.117 & 0.020 & 0.370 & 0.164 & 0.090 \\
TextFooler       & 0.100 & 0.052 & 0.000 & 0.260 & 0.091 & 0.091 & 0.210 & 0.137 & 0.010 & 0.340 & 0.170 & 0.120 \\
BertAttack       & 0.130 & 0.067 & 0.000 & 0.250 & 0.097 & 0.131 & 0.160 & 0.096 & 0.010 & 0.300 & 0.132 & 0.160 \\
TrivInsert       & 0.140 & 0.054 & 0.030 & 0.210 & 0.072 & 0.081 & 0.290 & 0.198 & 0.020 & 0.380 & 0.220 & 0.180 \\
TextBugger       & 0.120 & 0.062 & 0.020 & 0.190 & 0.080 & 0.071 & 0.240 & 0.164 & 0.000 & 0.350 & 0.191 & 0.100 \\
DrunkAgent       & 0.150 & 0.085 & 0.130 & 0.230 & 0.105 & 0.380 & 0.180 & 0.124 & 0.040 & 0.290 & 0.150 & 0.150 \\
\method~         & 0.130 & 0.086 & \underline{\textbf{0.182}} & 0.230 & 0.105 & \underline{\textbf{0.545}} & 0.225 & 0.139 & \underline{\textbf{0.113}} & 0.365 & 0.175 & \underline{\textbf{0.380}} \\
\bottomrule
\end{tabular}
}
\vspace{-5mm}
\end{table*}
\cref{tab:main_results} reports the attack effectiveness of all methods on three victim systems across two datasets. We can find:

\noindent\textbf{\method~achieves the highest exposure rate across all settings.}
On every combination of victim system and dataset, \method~obtains the highest E@10 and E@20.
The advantage is especially pronounced at $K{=}20$: on AgentRAG with CDs~\&~Vinyl, \method~achieves $\text{E@20}{=}0.545$, meaning the target item appears in the top-20 recommendations for over half of all benign users.
Across all six victim-dataset configurations, \method~achieves an average E@20 of $0.384$, compared with $0.199$ for the strongest individual baseline in each configuration.
\revised{This indicates that jointly exploiting both the semantic and structural attack surfaces, as designed in Sec.\ref{sec:method}, is substantially more effective than addressing either surface in isolation.}

\noindent\textbf{The joint semantic-structural design is essential for exploiting collaborative reflection.}
Interaction-level data poisoning methods (RandAttack, PopAttack, ExpPromot) inject fake trajectories but leave the item profile unchanged, achieving consistently low exposure rates with E@10 rarely exceeding $0.030$.
Text-level perturbation methods modify the profile but do not inject behavioural evidence, and their exposure rates are similarly low.
DrunkAgent, which rewrites the description more aggressively through surrogate-based optimisation, achieves moderate E@K on AgentRAG ($\text{E@20}{=}0.380$ on CDs~\&~Vinyl) but drops to near-zero on the other two paradigms.
These results corroborate the two-dimensional amplification analysis in Sec.\ref{sec:vuln_analysis}: influencing collaborative reflection requires evidence that is both semantically compatible with the reflection operator and topologically positioned to propagate across agents.
Methods that supply only one dimension cannot reliably enter and persist in the reflection-writeback cycle.
\method~is the only method that addresses both dimensions simultaneously, which explains its consistent advantage across all recommendation paradigms.

\noindent\textbf{\method~generalises across recommendation paradigms with varying memory surfaces.}
\method~achieves strong promotion across all three victim systems despite their different inference mechanisms.
On CDs~\&~Vinyl, E@20 ranges from $0.374$ (AgentCF) to $0.545$ (AgentRAG), and on Movies~\&~TV from $0.240$ (AgentCF) to $0.380$ (AgentRAG).
This cross-paradigm consistency arises because \method's semantic and structural components target the shared collaborative reflection mechanism rather than a particular inference pathway.
The variation in E@K across systems reflects the different amounts of memory context available during inference: systems that supply richer context (AgentRAG with retrieved preferences, AgentSEQ with historical sequences) provide more entry points for motif-aligned evidence to influence ranking.

\subsection{Stealthiness Analysis (RQ2)}
\label{sec:exp_stealthiness}

\revised{We evaluate stealthiness from four complementary aspects: recommendation quality preservation, text naturalness, content fidelity, and human evaluation.}

\noindent\textbf{Recommendation quality preservation.}
As shown in \cref{tab:main_results}, \method's H@K and N@K remain within a narrow range of the NoAttack baseline across all settings.

For example, on AgentCF with CDs~\&~Vinyl, H@20 is $0.250$ under \method~versus $0.260$ without attack, and N@20 is $0.083$ versus $0.082$.
\revised{The limited change in H@K and N@K indicates that target promotion does
not substantially disrupt aggregate recommendation performance.
From the platform operator's perspective, these utility metrics provide
little indication of anomalous behaviour, making the attack harder to
identify through routine performance monitoring.}

\begin{table}[ht]
\centering
\small
\setlength{\tabcolsep}{7pt}
\renewcommand{\arraystretch}{1.15}
\vspace{-3mm}
\caption{Attack imperceptibility in terms of perplexity.}
\label{tab:ppl}
\resizebox{\columnwidth}{!}{%
\begin{tabular}{lcccc}
\toprule
\multirow{2}{*}{\textbf{Dataset}} & \multicolumn{2}{c}{\textbf{NoAttack}} & \multirow{2}{*}{\textbf{DrunkAgent}} & \multirow{2}{*}{\textbf{Ours}} \\
\cmidrule(lr){2-3}
 & \textbf{Range} & \textbf{Mean} & & \\
\midrule
CDs \& Vinyl  & {\footnotesize [10.31, 90.02]}  & 28.78 & 49.93 & \textbf{32.22} \\
Movies \& TV  & {\footnotesize [12.91, 374.28]} & 72.85 & 50.65 & \textbf{39.35} \\
\bottomrule
\end{tabular}%
}
\vspace{-3mm}
\end{table}

\noindent\textbf{Text naturalness (PPL).}
\cref{tab:ppl} reports the perplexity of modified profiles alongside the original profiles' mean and range.
On CDs~\&~Vinyl, \method~achieves a PPL of $32.22$, close to the original mean of $28.78$ and well within the original range $[10.31,90.02]$, while DrunkAgent's PPL rises to $49.93$.
On Movies~\&~TV, the original profiles exhibit high variance (range $[12.91,374.28]$, mean $72.85$), reflecting heterogeneous description quality across items in this domain.
\method's PPL of $39.35$ falls comfortably within this range. 
In both domains, the perplexity of \method's profile falls within the range observed for ordinary item descriptions.

\begin{table}[ht]
\centering
\small
\setlength{\tabcolsep}{7pt} 
\vspace{-3mm}
\caption{Attack imperceptibility in terms of ROUGE.}
\label{tab:rouge}
\resizebox{1\columnwidth}{!}{%
\begin{tabular}{lcccccc}
\toprule
\multirow{2}{*}{\textbf{Attacks}} & \multicolumn{3}{c}{\textbf{CDs \& Vinyl}} & \multicolumn{3}{c}{\textbf{Movies \& TV}} \\ 
\cmidrule(lr){2-4}\cmidrule(lr){5-7}
 & \textbf{@1} & \textbf{@2} & \textbf{@L} & \textbf{@1} & \textbf{@2} & \textbf{@L} \\ 
\midrule
TextFooler  & 0.200 & 0.000 & 0.200 & 0.251 & 0.000 & 0.121 \\
DrunkAgent  & 0.294 & 0.000 & 0.100 & 0.175 & 0.000 & 0.087 \\
\method~    & \textbf{0.321} & \textbf{0.109} & \textbf{0.268} & \textbf{0.294} & \textbf{0.090} & \textbf{0.157} \\
\bottomrule
\end{tabular}%
\vspace{-3mm}
}
\end{table}

\noindent\textbf{Content fidelity (ROUGE).}
\cref{tab:rouge} compares the lexical overlap between modified and original profiles.
\method~achieves the highest ROUGE-1, ROUGE-2, and ROUGE-L scores across both datasets.
Notably, \method~is the only method with non-zero ROUGE-2 scores ($0.109$ on CDs~\&~Vinyl, $0.090$ on Movies~\&~TV), indicating that it preserves bigram-level phrasing from the original profile.
This is a direct consequence of the motif-aware contrastive rewriting design, which anchors the rewritten profile to the original content while incorporating transferable preference motifs.
TextFooler and DrunkAgent achieve zero ROUGE-2 across both datasets, indicating that their modifications disrupt the profile's local phrasing structure even when word-level overlap (ROUGE-1) is maintained.

\noindent\revised{\textbf{Human evaluation.} 
Since item profiles are ultimately read by shoppers, we further conduct a human evaluation with 20 participants, each assessing 120 descriptions from CDs \& Vinyl in randomised order, 40 each from original profiles and those rewritten by DrunkAgent and \method.
As Table~\ref{tab:human} shows, \method~ profiles are flagged as manipulated in 21.9\% of cases, close to 19.3\% for original profiles and lower than 28.7\% for DrunkAgent. Their factual consistency is also close to the original profiles (4.16 vs. 4.31 on a 1--5 scale), with no significant difference between the two ($p=0.25$, paired $t$-test), whereas DrunkAgent obtains 4.02. These results complement the automatic metrics and show that \method~remains difficult to distinguish from genuine profiles while largely preserving their factual content.}
\begin{table}[t]
    \centering
    \caption{\revised{Attack imperceptibility in terms of human evaluation (20 participants, 120 descriptions each).}}
    \label{tab:human}
    \setlength{\tabcolsep}{4pt}
    \renewcommand{\arraystretch}{1.1}
    \begin{tabular}{lcc}
    \toprule
    \textbf{Attacks} & \textbf{Flagged as Manipulated} $\downarrow$ & \textbf{Factual Consistency} $\uparrow$ \\
    \midrule
    Original              & 19.3\% & $4.31 \pm 0.47$ \\
    DrunkAgent            & 28.7\% & $4.02 \pm 0.95$ \\
    \textit{VirusCascade} & 21.9\% & $4.16 \pm 0.62$ \\
    \bottomrule
    \end{tabular}
    \vspace{-7pt}
\end{table}

\subsection{Ablation Study (RQ3)}
\begin{table*}[ht]
\centering
\vspace{-3mm}
\small
\caption{Ablation study: effect of removing semantic injection (w.o SemInj) or structural injection (w.o StruInj).}
\label{tab:ablation_study}
\definecolor{ERPurple}{RGB}{242,235,255}
\resizebox{0.95\textwidth}{!}{%
\begin{tabular}{lcc>{\columncolor{blue!8}}c cc>{\columncolor{blue!8}}c cc>{\columncolor{blue!8}}c cc>{\columncolor{blue!8}}c}
\toprule
\rowcolor{gray!15}
\textit{Victim RS} & \multicolumn{12}{c}{\textit{AgentCF}} \\ 
\cmidrule{1-13}
\multirow{2}{*}{\textbf{Variant}} & \multicolumn{6}{c}{\textbf{CDs \& Vinyl}} & \multicolumn{6}{c}{\textbf{Movies \& TV}} \\
\cmidrule(lr){2-7}\cmidrule(lr){8-13}
 & \textbf{H@10} & \textbf{N@10} & \textbf{E@10} & \textbf{H@20} & \textbf{N@20} & \textbf{E@20} & \textbf{H@10} & \textbf{N@10} & \textbf{E@10} & \textbf{H@20} & \textbf{N@20} & \textbf{E@20} \\ \hline
NoAttack         & 0.130 & 0.050 & (-)   & 0.260 & 0.082 & (-)   & 0.220 & 0.125 & (-)   & 0.310 & 0.148 & (-)   \\
w.o SemInj   & 0.110 & 0.058 & 0.010 & 0.250 & 0.094 & 0.071 & 0.200 & 0.135 & 0.020 & 0.320 & 0.164 & 0.170 \\
w.o StruInj  & 0.100 & 0.046 & 0.020 & 0.260 & 0.086 & 0.131 & 0.160 & 0.101 & 0.000 & 0.300 & 0.135 & 0.080 \\
\method~         & 0.120 & 0.057 & \underline{\textbf{0.111}} & 0.250 & 0.083 & \underline{\textbf{0.374}} & 0.205 & 0.123 & \underline{\textbf{0.070}} & 0.305 & 0.138 & \underline{\textbf{0.240}} \\
\hline 

\midrule
\rowcolor{gray!15}
\textit{Victim RS} & \multicolumn{12}{c}{\textit{AgentSEQ}} \\ 
\cmidrule(lr){2-7}\cmidrule(lr){8-13}
\textbf{Variant} & \textbf{H@10} & \textbf{N@10} & \textbf{E@10} & \textbf{H@20} & \textbf{N@20} & \textbf{E@20} & \textbf{H@10} & \textbf{N@10} & \textbf{E@10} & \textbf{H@20} & \textbf{N@20} & \textbf{E@20} \\ \hline
NoAttack         & 0.130 & 0.056 & (-)   & 0.270 & 0.091 & (-)   & 0.240 & 0.126 & (-)   & 0.350 & 0.154 & (-)   \\
w.o SemInj   & 0.110 & 0.065 & 0.010 & 0.250 & 0.100 & 0.071 & 0.180 & 0.127 & 0.050 & 0.300 & 0.157 & 0.230 \\
w.o StruInj  & 0.100 & 0.058 & 0.040 & 0.260 & 0.097 & 0.202 & 0.140 & 0.086 & 0.000 & 0.280 & 0.120 & 0.040 \\
\method~         & 0.120 & 0.053 & \underline{\textbf{0.131}} & 0.240 & 0.087 & \underline{\textbf{0.465}} & 0.220 & 0.117 & \underline{\textbf{0.090}} & 0.330 & 0.146 & \underline{\textbf{0.300}} \\
\hline 

\midrule
\rowcolor{gray!15}
\textit{Victim RS} & \multicolumn{12}{c}{\textit{AgentRAG}} \\ 
\cmidrule(lr){2-7}\cmidrule(lr){8-13}
\textbf{Variant} & \textbf{H@10} & \textbf{N@10} & \textbf{E@10} & \textbf{H@20} & \textbf{N@20} & \textbf{E@20} & \textbf{H@10} & \textbf{N@10} & \textbf{E@10} & \textbf{H@20} & \textbf{N@20} & \textbf{E@20} \\ \hline
NoAttack         & 0.140 & 0.056 & (-)   & 0.250 & 0.083 & (-)   & 0.240 & 0.155 & (-)   & 0.380 & 0.189 & (-)   \\
w.o SemInj   & 0.130 & 0.078 & 0.020 & 0.420 & 0.147 & 0.101 & 0.240 & 0.158 & 0.090 & 0.370 & 0.190 & 0.220 \\
w.o StruInj  & 0.130 & 0.070 & 0.020 & 0.290 & 0.111 & 0.141 & 0.190 & 0.128 & 0.010 & 0.300 & 0.156 & 0.110 \\
\method~         & 0.130 & 0.086 & \underline{\textbf{0.182}} & 0.230 & 0.105 & \underline{\textbf{0.545}} & 0.225 & 0.139 & \underline{\textbf{0.113}} & 0.365 & 0.175 & \underline{\textbf{0.380}} \\
\bottomrule
\end{tabular}
}
\end{table*}

\begin{figure*}[t]
    \centering
    \vspace{-3mm}
    \includegraphics[width=0.95\linewidth]{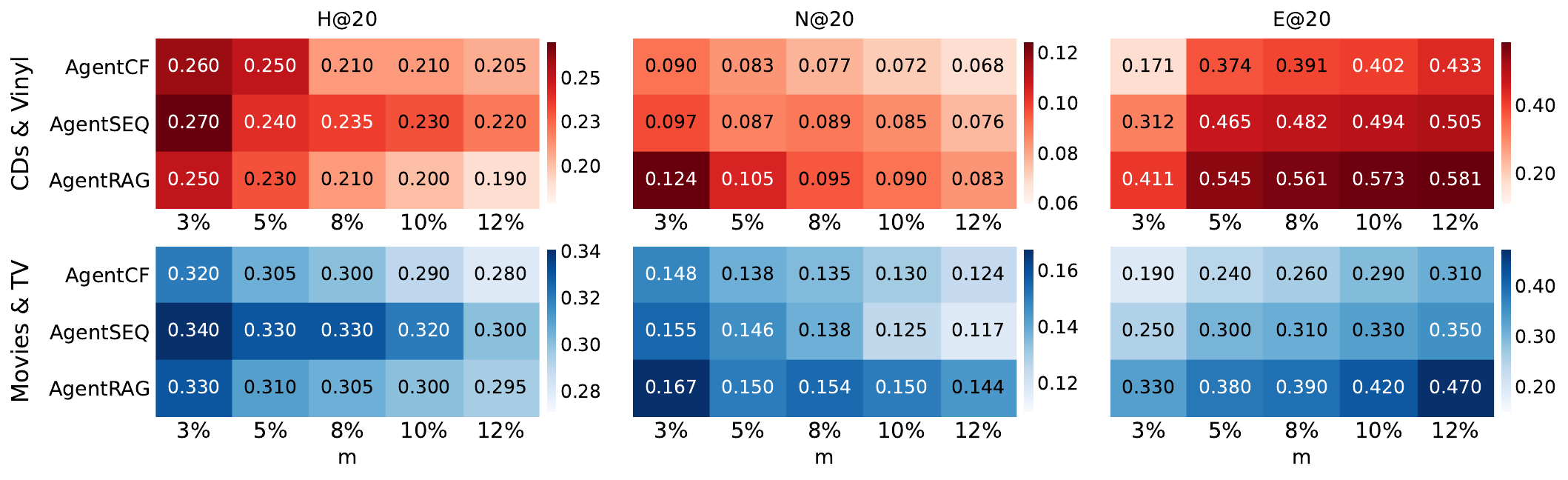}
    \vspace{-3mm}
    \caption{Effect of malicious user fraction $m$.}
    \label{fig:sens_m}
    \vspace{-5mm}
\end{figure*}

\noindent\revised{To disentangle the contributions of the two adversary-controlled components, we evaluate two variants. \textbf{w.o SemInj} retains the designed trajectories but uses the original item profile, isolating structural injection. \textbf{w.o StruInj} uses the motif-aware rewritten profile but replaces the designed trajectories with random interactions, isolating semantic injection. Table~\ref{tab:ablation_study} reports results across all three victim systems. We find:
}

\noindent\textbf{Both components are necessary and complementary.}
Removing either component consistently reduces E@K.
On AgentCF with CDs~\&~Vinyl, the full method achieves $\text{E@20}{=}0.374$, while removing semantic injection reduces it to $0.071$ and removing structural injection reduces it to $0.131$.
The pattern is consistent across all victim systems, confirming that the two components contribute non-redundant attack surfaces.

\noindent\textbf{Structural injection provides the propagation backbone.}
\revised{Across most configurations, removing structural injection causes a larger E@K drop than removing semantic injection. This is consistent with the cross-agent propagation analysis: without trajectories routed through popular anchor items, injected evidence is less likely to enter reflection contexts shared by benign users. The designed trajectories thus provide topological pathways for motif-aligned signals to reach broader user groups. This effect is most visible on AgentSEQ with Movies \& TV, where removing structural injection reduces E@20 from 0.300 to 0.040, a 7.5$\times$ reduction. 
}

\noindent\textbf{Semantic injection enables admission and persistence.}
Complementing the structural propagation path, the motif-aware profile ensures that the target item is interpreted as preference-compatible during reflection.
The variant \emph{w.o SemInj} yields substantially lower E@K because the unmodified profile lacks transferable preference motifs needed to pass the reflection operator's consistency check.
This aligns with the persistence analysis in Sec.\ref{sec:vuln:persistence}, where local compatibility was identified as a necessary condition for probe admission.

\noindent\textbf{The two components exhibit synergistic interaction.}
\revised{The full method consistently outperforms either ablated variant, showing that semantic and structural injection provide complementary benefits.}
Semantic injection makes the target easier to admit and persist within the reflection-writeback cycle, while structural injection places the target in high-connectivity contexts where the admitted evidence can propagate to non-injected agents.
The synergy arises because the two components exploit the two orthogonal dimensions of the amplification surface identified in Sec.\ref{sec:vuln_analysis}: temporal persistence and topological propagation.

\subsection{Sensitivity Analysis (RQ4)}
\label{sec:exp_sensitivity}

We examine how attack performance varies with two key hyperparameters: the malicious user fraction~$m$ and the semantic-behavioural trade-off coefficient~$\alpha$ in the dual-cost transition graph (Sec.\ref{sec:method_structural_injection}).

\noindent\textbf{Effect of the malicious user fraction ($m$).}
\cref{fig:sens_m} reports E@20, H@20, and N@20 as $m$ varies in $\{3\%, 5\%, 8\%, 10\%, 12\%\}$ across all three victim systems on both datasets.
Two findings are consistent across all settings.

First, E@20 increases monotonically with $m$ but exhibits clear diminishing returns beyond $m{=}5\%$.
The largest marginal gain occurs between $m{=}3\%$ and $m{=}5\%$.
On AgentRAG with CDs~\&~Vinyl, E@20 jumps from $0.411$ to $0.545$ ($+0.134$), while the entire range from $5\%$ to $12\%$ yields only an additional $+0.036$.
Similar patterns appear on AgentSEQ ($+0.153$ versus $+0.040$) and AgentCF ($+0.203$ versus $+0.059$).
Even at the smallest budget $m{=}3\%$, \method~already achieves E@20 of $0.171$ (AgentCF), $0.312$ (AgentSEQ), and $0.411$ (AgentRAG) on CDs~\&~Vinyl, all exceeding the strongest baseline at the default $m{=}5\%$.
This confirms that \method's effectiveness stems from the quality of its semantic and structural design rather than from the sheer volume of fake users, and demonstrates the efficiency of the anchor-guided trajectory routing in propagating injected evidence with a minimal attacker footprint.

Second, recommendation quality (H@20 and N@20) decreases gradually as $m$ grows.
On AgentCF with CDs~\&~Vinyl, H@20 declines from $0.260$ at $m{=}3\%$ to $0.205$ at $m{=}12\%$, while N@20 drops from $0.090$ to $0.068$.
The degradation is moderate and approximately linear in $m$, presenting the attacker with a smooth trade-off between promotion strength and recommendation fidelity.
At the default $m{=}5\%$, this trade-off is favourable: E@20 has already captured the majority of the achievable gain while recommendation quality remains close to the no-attack baseline.

\begin{figure*}[t]
    \centering
    \vspace{-7mm}
    \includegraphics[width=0.90\linewidth]{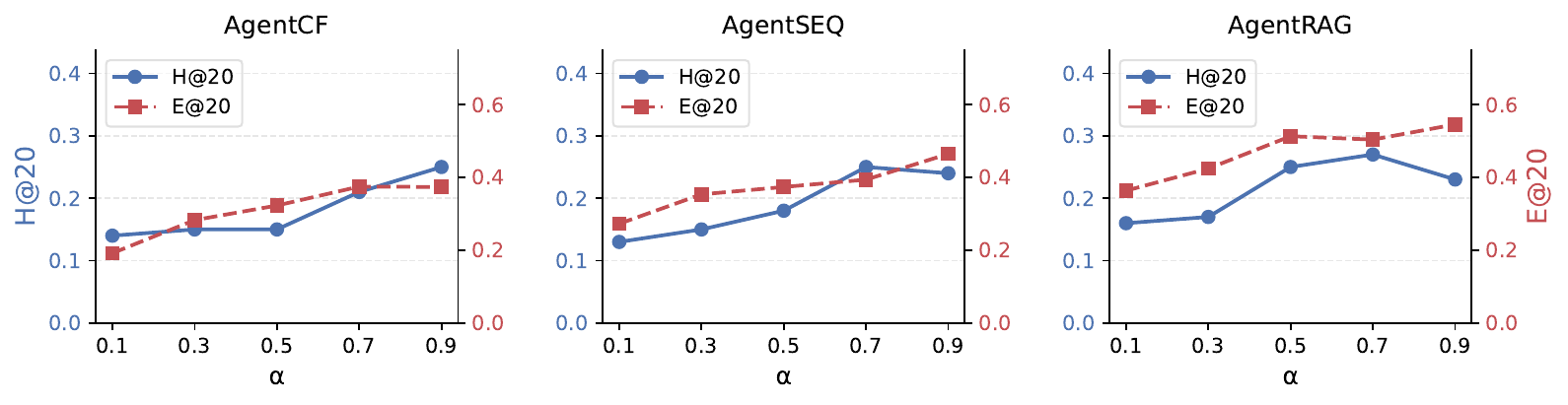}
    \vspace{-3mm}
    \caption{Effect of the coefficient $\alpha$.}
    \vspace{-5mm}
    \label{fig:sens_alpha}
    
\end{figure*}

\noindent\textbf{Effect of the trade-off coefficient ($\alpha$).}
The coefficient $\alpha$ controls the relative weight of semantic similarity versus behavioural naturalness when routing trajectories through the transition graph.
\cref{fig:sens_alpha} reports H@20 and E@20 as $\alpha$ varies in $\{0.1, 0.3, 0.5, 0.7, 0.9\}$ on CDs~\&~Vinyl.

Across all three victim systems, low values of $\alpha$ lead to substantially lower E@20.
On AgentCF, E@20 rises from $0.192$ at $\alpha{=}0.1$ to $0.375$ at $\alpha{=}0.7$, where it stabilises ($0.374$ at $\alpha{=}0.9$).
This indicates that semantic coherence in trajectory routing plays a more critical role than behavioural naturalness for attack effectiveness: when $\alpha$ is too low, the selected trajectories pass through items that are behaviourally adjacent but semantically distant from the target, weakening the motif alignment that drives reflection admission.
The optimal $\alpha$ is system-dependent but consistently favours the upper range.
AgentCF saturates around $\alpha{=}0.7$, while AgentSEQ and AgentRAG continue to improve up to $\alpha{=}0.9$. 
These two systems supply richer contextual evidence during inference, whether through historical interactions or retrieved preferences, creating a reflection environment that rewards semantically coherent trajectories even more.
On AgentRAG, E@20 increases steadily from $0.364$ at $\alpha{=}0.1$ to $0.513$ at $\alpha{=}0.5$, then stabilises around $0.50$--$0.55$ before reaching its peak at $\alpha{=}0.9$.
This suggests that once trajectories achieve sufficient semantic alignment with the target, further increases in $\alpha$ yield marginal gains until a threshold is crossed where the retrieval mechanism consistently surfaces motif-aligned evidence.

\noindent\subsection{\revised{Scalability Analysis (RQ5)}}
\label{ssec:scalability_analysis}
\revised{We further evaluate \method~beyond the default experimental scale
along two dimensions: active-agent population and the number of
interaction rounds.
For population scaling, we fix the attacker budget at five users while
increasing the active population from 100 to 500, reducing the effective
malicious-user fraction from 5\% to 1\%.
Although E@20 attenuates with population size, \method~remains above
DrunkAgent across all three victim systems. 
For interaction-round scaling, we extend to 20 rounds
without additional adversarial interactions.
The promotion effect gradually attenuates but remains measurable, while H@20 shows no progressive degradation.
Detailed results are provided in Appendices~\ref{ssec:fixed_maliUser} and~\ref{ssec:more_round}.
}

\subsection{Transferability Analysis \revised{(RQ6)}}
\label{sec:exp_transferability}

\begin{table*}[ht]
\centering
\small
\vspace{-3mm}
\caption{Attack transferability with different auxiliary models $\mathcal{M}_{\mathrm{aux}}$.}
\label{tab:transfer_maux}
\definecolor{ERPurple}{RGB}{242,235,255}  
\setlength{\tabcolsep}{6pt}
\begin{tabular}{llcc>{\columncolor{blue!8}}c cc>{\columncolor{blue!8}}c cc>{\columncolor{blue!8}}c}
\toprule
\multirow{2}{*}{\textbf{Dataset}} & \multirow{2}{*}{\textbf{Victim RS}} 
& \multicolumn{3}{c}{\textbf{LLaMA-3}} 
& \multicolumn{3}{c}{\textbf{GPT-4o}} 
& \multicolumn{3}{c}{\textbf{Gemini-2.5}} \\
\cmidrule(lr){3-5}\cmidrule(lr){6-8}\cmidrule(lr){9-11}
 & & \textbf{H@20} & \textbf{N@20} & \textbf{E@20} & \textbf{H@20} & \textbf{N@20} & \textbf{E@20} & \textbf{H@20} & \textbf{N@20} & \textbf{E@20} \\
\cmidrule{1-11}

\multirow{3}{*}{\textbf{CDs \& Vinyl}} 
 & AgentCF  & 0.320 & 0.107 & 0.151 & 0.250 & 0.083 & 0.374 & 0.340 & 0.119 & 0.333 \\
 & AgentSEQ & 0.410 & 0.129 & 0.160 & 0.240 & 0.087 & 0.465 & 0.330 & 0.112 & 0.283 \\
 & AgentRAG & 0.320 & 0.096 & 0.170 & 0.230 & 0.105 & 0.545 & 0.350 & 0.115 & 0.364 \\
\midrule

\multirow{3}{*}{\textbf{Movies \& TV}} 
 & AgentCF  & 0.250 & 0.102 & 0.140 & 0.290 & 0.134 & 0.240 & 0.340 & 0.161 & 0.140 \\
 & AgentSEQ & 0.330 & 0.144 & 0.150 & 0.300 & 0.142 & 0.300 & 0.340 & 0.165 & 0.210 \\
 & AgentRAG & 0.300 & 0.148 & 0.160 & 0.310 & 0.150 & 0.380 & 0.330 & 0.175 & 0.270 \\
 \bottomrule
\end{tabular}%
\vspace{-3mm}
\end{table*}

\revised{A practical attack should remain effective beyond its default configuration without sacrificing recommendation utility or profile stealthiness.
We evaluate transferability along two dimensions: varying the auxiliary model $\mathcal{M}_{\mathrm{aux}}$ used for semantic injection and simplifying the victim from a dual-agent to a user-only architecture.}

\noindent\revised{\textbf{Transferability across auxiliary models ($\mathcal{M}_{\mathrm{aux}}$)}.
To evaluate whether this alignment is necessary, we evaluate the attack using three auxiliary models: \texttt{LLaMA-3}, \texttt{GPT-4o}, and \texttt{Gemini-2.5}. 
In each case, only $\mathcal{M}_{\mathrm{aux}}$ is changed while all other attack parameters and the victim system remain identical.
As shown in \cref{tab:transfer_maux}, \method~achieves positive E@20 with all auxiliary models, while H@20 and N@20 remain broadly comparable across the corresponding settings.
GPT-4o yields the strongest promotion in most configurations, whereas Gemini-2.5 and LLaMA-3 still provide non-trivial exposure gains, showing that \method~does not rely on model-specific alignment between $\mathcal{M}_{\mathrm{aux}}$ and the victim backbone.}

\revised{Since $\mathcal{M}_{\mathrm{aux}}$ also determines the target profiles generated during semantic injection, we further evaluate their stealthiness in \cref{tab:aux_imperceptibility}.
Across all models, the resulting profiles maintain text naturalness, with PPL remaining within the corresponding NoAttack ranges, while also preserving content fidelity under ROUGE. This indicates that cross-model transfer does not require conspicuous profile modifications.}

\begin{table}[t]
\centering
\caption{\revised{Attack imperceptibility with different auxiliary models
$\mathcal{M}_{\mathrm{aux}}$. NoAttack PPL is reported as
[min, max].} 
}
\label{tab:aux_imperceptibility}
\resizebox{\columnwidth}{!}{
\setlength{\tabcolsep}{3pt}
\begin{tabular}{lllcccc}
\toprule
\multirow{2}{*}{\textbf{Dataset}} &
\multirow{2}{*}{\textbf{$\mathcal{M}_{\mathrm{aux}}$}} &
\multicolumn{2}{c}{\textbf{PPL}} &
\multicolumn{3}{c}{\textbf{ROUGE}} \\
\cmidrule(lr){3-4}
\cmidrule(lr){5-7}
& &
\textbf{NoAttack} &
\textbf{Ours} &
\textbf{@1} &
\textbf{@2} &
\textbf{@L} \\
\midrule

\multirow{3}{*}{CDs \& Vinyl}
& LLaMA-3     & [21.64, 107.62]    & 56.28    & 0.267    & 0.103    & 0.267 \\

& GPT-4o      & [10.31, 90.02]    & 32.22    & 0.321     & 0.109    & 0.268 \\

& Gemini-2.5  & [11.78, 113.24]   & 31.82    & 0.341     & 0.125    & 0.268 \\
\midrule

\multirow{3}{*}{Movies \& TV}
& LLaMA-3     & [13.75, 409.59]    & 76.20    & 0.412    & 0.091    & 0.206 \\

& GPT-4o      & [12.91, 374.28]    & 39.35    & 0.294    & 0.090    & 0.157 \\

& Gemini-2.5  & [13.62, 341.82]    & 30.76    & 0.320    & 0.048    & 0.180 \\

\bottomrule
\end{tabular}}
\vspace{-3mm}
\end{table}

\noindent\textbf{Transferability to user-only agent architectures.}
Many practical LLM-ARS adopt a simpler user-only agent design where items are represented by static profiles rather than autonomous agents.
In such systems, item-side memory is not updated during reflection, which removes one of the two propagation channels exploited by \method.
To evaluate whether \method~remains effective under this simplified architecture, we re-instantiate all three victim systems as user-only variants and compare \method~with the strongest baselines (TextBugger, TextFooler, DrunkAgent).
\revised{Due to space limits, we report the detailed results in Appendix~\ref{ssec:transfer_useronly}. 
\method~consistently achieves the highest E@K across all user-only configurations while maintaining comparable H@K and N@K.
For example, on CDs~\&~Vinyl, \method~obtains E@20 of $0.152$ (AgentCF), $0.283$ (AgentSEQ), and $0.202$ (AgentRAG), outperforming the best baseline in each configuration by $1.5\times$ to $2.5\times$.
Although exposure generally decreases relative to the dual-agent setting, the attack remains effective after removing item-side memory updates, showing that its effectiveness does not depend exclusively on
item-side memory propagation.}
\revised{Since profile generation depends only on semantic injection through $\mathcal{M}_{\mathrm{aux}}$, changing the victim to a user-only architecture leaves the generated profiles unchanged. Its PPL and ROUGE therefore remain identical to the GPT-4o setting in \cref{tab:aux_imperceptibility}.}

\section{Background and Related Work}
\label{sec:RelatedWork}

\subsection{LLM-based Agentic Recommender System}
Recent advances in large language models (LLMs) are transforming recommender systems (RS) from static prediction pipelines into LLM-based agentic recommender systems (LLM-ARS)~\cite{shang2026agentrecbench}.
Unlike conventional models that optimise fixed scoring functions over historical interactions, LLM-ARS equip agents with natural-language memory and reflection capabilities~\cite{yu2025intelligent, wang2024macrec,zhang2026llms, liu2025agentcf++}, enabling them to synthesise contextual knowledge, revise preference states, and adapt decisions across multiple interaction rounds.
Early work focuses on single-agent user modelling~\cite{yu2025intelligent}, where an autonomous user agent perceives interaction feedback and dynamically updates its preferences, improving recommendation accuracy over static baselines.
This paradigm has since been extended to multi-agent collaborative architectures in which both user and item agents interact, reason, and update their states jointly~\cite{zhang2024agentcf, wang2024macrec, zhang2026llms,liu2025agentcf++}.
AgentCF~\cite{zhang2024agentcf}, for example, instantiates users and items as dual LLM agents that collaboratively reflect on interaction outcomes to iteratively refine their memories.
ARAG~\cite{maragheh2025arag} further introduces a multi-agent RAG architecture for E-commerce recommendation, demonstrating the practical value of agentic collaboration in capturing complex user intent.

\subsection{Attacks on LLM-based Recommender Systems}
Existing work can be organised along a progression of attack surfaces.
The first line targets \emph{interaction-level data poisoning} in conventional RS~\cite{wang2024unveiling,nazary2025stealthy}, injecting fabricated ratings or interactions to bias the training of static scoring models.
As LLMs are integrated into RS and decision-making shifts toward natural-language reasoning, a second line explores \emph{text-level adversarial perturbations}~\cite{guan2026c3h,ning2024cheatagent,zhao2025lance,gu2026llm,guan2026venomrec}, applying imperceptible edits to item descriptions, reviews, or prompt templates to mislead the LLM's semantic understanding.
AgentSA~\cite{gu2026llm}, for instance, leverages LLM agents to autonomously craft user profiles and reviews that maximise manipulation impact while maintaining behavioural stealthiness.
Despite their effectiveness, both lines assume a \emph{static} recommendation pipeline, whether fixed training data or single-turn inference, and therefore cannot exploit the recurrent memory dynamics of LLM-ARS.
A third line begins to address this gap.
DrunkAgent~\cite{yang2025drunkagent} achieves memory corruption by covertly tampering with the external memory of an LLM-ARS agent, continuously misleading its decisions across multi-round reasoning.
However, existing attacks focus on isolated memory contamination but overlook a more dangerous threat: system-level amplification risks caused by collaborative reflection and multi-agent reasoning in advanced LLM-ARS.

\section{Conclusion and Future Work}
In this paper, we identified \emph{collaborative-reflection hijacking}, a systemic vulnerability in LLM-ARS where the collaborative reflection mechanism simultaneously enables adversarial evidence to be laundered into legitimate preferences and propagated across agents through shared interaction contexts.
\revised{A controlled vulnerability analysis established evidence for two exploitable properties underlying this threat: \emph{reflective persistence} and \emph{cross-agent propagation}.}
Building on these findings, we proposed \method, the first black-box targeted promotion attack that jointly shapes semantic and structural attack surfaces to exploit both properties, with the victim system's own reflection driving amplification without further adversarial intervention.
\revised{Extensive experiments across four real-world datasets and three victim architectures demonstrate state-of-the-art effectiveness and practical stealthiness under the evaluated black-box constraints.}
This work offers both an effective attack framework and a lens to understand how multi-agent reflection creates amplification pathways beyond the reach of conventional poisoning, providing guidance for building more robust agentic recommendation architectures.

\section*{Acknowledgment}
We thank the anonymous reviewers for their constructive feedback. 
This research is supported by the Ministry of Education, Singapore, under its Academic Research Fund Tier 2 (Award MOE-T2EP20125-0005).

\section*{Ethics Considerations}
This work studies a security vulnerability in LLM-powered agentic recommender systems (LLM-ARS) under controlled conditions. All experiments were conducted in a local, sandboxed simulation using publicly available Amazon review datasets; no real-world platform, production API, or real user was targeted, and all attacker-controlled users were synthetic. No personal data beyond the publicly released anonymised corpus was used. In the human study of Sec.~V-C, informed volunteers evaluated public item descriptions offline without interacting with the recommender system. We disclosed our findings to the developers of the affected open-source LLM-ARS frameworks prior to publication. Our goal is to expose an overlooked attack surface in collaborative reflection and support the development of more robust agentic recommender systems and principled countermeasures.



%
\bibliographystyle{IEEEtran}
\bibliography{refs}

\appendix
\makeatletter
\renewcommand{\thesection}{\Alph{section}} 
\renewcommand{\@seccntformat}[1]{\csname the#1\endcsname.\quad} 
\makeatother
\section{Appendix}
\label{sec:appendix}

\subsection{More Setup for Vulnerability Analysis}
\label{app:vuln_setup}
This section provides the full experimental protocol for the two vulnerability probes presented in~Sec.\ref{sec:vuln_analysis}.
The probes are designed to explore two structural properties of collaborative reflection, namely reflective persistence (Sec.\ref{sec:vuln:persistence}) and cross-agent propagation (Sec.\ref{sec:vuln:propagation}). 
Both probes are conducted on AgentCF~\cite{zhang2024agentcf} with the CDs~\&~Vinyl dataset under the leave-one-out sequential~(LS) evaluation protocol.
The dataset contains 100 benign users and their associated items.

\noindent\textbf{Hypothesis 1: Reflective Persistence.}
We construct a single synthetic probe user and append it to a temporary copy of the CDs~\&~Vinyl training split.
The probe user's initial memory profile is seeded with four rare probe phrases whose natural occurrence rate in the corpus is below 5\%:
\emph{``complete after midnight session''}, \emph{``emotionally charged''}, \emph{``unforgettable performances''} and \emph{``genuine musical experiences.''}
Its interaction history is set to $[i_{26},\, i_{117},\, i_{945},\, i_{121},\, i_{125}]$, where $i_{945}$ (``Complete After Midnight Session'') is positioned so that it falls within the training split under LS evaluation.
AgentCF is run for 5 interaction-reflection rounds.
The probe user participates only in round~1 and is removed before round~2, leaving 4 purely benign rounds in which the system evolves without any probe input.
A parallel \emph{no-probe control arm} runs the same configuration on the original (unmodified) training split, ensuring that all 100 real users receive identical benign traffic across both runs.
Memory snapshots of all agents (100 real users plus items) are saved after each round.

For each real user, we record the following quantities.
\emph{Admission} is determined by checking whether the user's short-term memory $M_u^s$ contains at least one of the four rare probe phrases.
Since AgentCF's reflection operator rewrites $M_u^s$ at each round rather than appending to it, the presence of a marker phrase indicates that reflection actively regenerated it.
The \emph{admission round} is the earliest round at which this criterion is first satisfied.
Subsequent measurements are aligned by \emph{lag}, where lag\,$k$ denotes the $k$-th round after the user's individual admission round.
The \emph{reuse rate} at lag\,$k$ is the fraction of admitted users (whose admission round is early enough to be observed at lag\,$k$) whose short-term memory still contains at least one marker phrase.
The \emph{elaboration rate} at lag $k$ is the fraction of such users whose
reflection trace preserves the probe claim and adds a new rationale,
implication, or preference explanation beyond verbatim repetition or
paraphrase, as judged by GPT-4o-mini with a binary classification prompt.
The \emph{baseline regeneration rate} is the frequency at which the same rare phrases appear in the no-probe control arm, providing the reference for all comparisons reported in Table~\ref{tab:persistence}.

\noindent\textbf{Hypothesis 2: Cross-Agent Propagation.}
We reuse the same synthetic probe user and AgentCF configuration described above.
While Hypothesis~1 tracks the fate of the probe claim within admitted users, Hypothesis~2 shifts focus to the \emph{non-injected} benign users and asks whether a single local memory update can produce measurable downstream effects.
After the 5 interaction-reflection rounds, we construct the user--item bipartite interaction graph from the training split and compute the shortest-path hop distance from the probe user~$u_0$ to every benign user.
Two users are connected through a path $u_0 \text{-} i \text{-} u$ if both have interacted with the same item~$i$.
Hop-1 denotes users sharing at least one item with~$u_0$.
Hop-2 denotes users reachable through exactly one intermediate user.
Hop${>}$2 groups all remaining users.
For each non-injected user, we measure two quantities: recommendation drift and claim adoption, both reported as excess over the no-probe control arm.

\noindent\revised{\textbf{Common-frequency markers}. As shown in
Table~\ref{tab:commom_marker}, common-frequency markers exhibit a
persistence pattern consistent with that observed for rare markers.
Despite a higher no-probe regeneration rate (15.0\% vs. 2.0\%),
their reuse rates remain between 68.7\% and 76.4\% across subsequent
rounds, substantially above the corresponding benign baseline.
}

\begin{table}[t]
\centering
\vspace{-1mm}
\caption{\revised{Reflective persistence under rare and
common-frequency diagnostic markers. Lag columns report reuse rate.}}
\vspace{-1mm}
\label{tab:commom_marker}
\begin{tabular}{lccccc}
\toprule
Marker
& No-Probe Regen.
& Lag-1
& Lag-2
& Lag-3
& Lag-4 \\
\midrule
Rare
& 2.0\%
& 78.0\%
& 81.7\%
& 79.3\%
& 81.6\%  \\
Common
& 15.0\%
& 72.0\%
& 73.4\%
& 68.7\%
& 76.4\%  \\
\bottomrule
\end{tabular}
\vspace{-8pt}
\end{table}
\noindent\revised{\textbf{Elaboration judgement.} Each admitted user's reflection trace at a given lag is submitted to GPT-4o-mini at temperature 0 with the binary prompt below, which supplies the four marker phrases, the trace, and the three conditions defined in Sec.~\ref{sec:vuln:persistence}, and requests a single label with no explanation.}
\begin{Prompt}{Elaboration Prompt}
\revised{You are annotating memory traces from a recommender agent.
A probe claim was previously seeded into some agents' memories.
The claim is expressed through these marker phrases: \{marker\_phrases\}.
Below is one agent's reflection trace from a later round: \{reflection\_trace\}.
Output 1 only if all hold:
\begin{itemize}
    \item the claim is preserved, not contradicted or dropped;
    \item the trace adds a rationale, implication, or preference
    explanation not stated by the marker phrases;
    \item the addition is not repetition or paraphrase.
\end{itemize}
If any condition fails, label 0. Output one character (1 or 0) only.}
\end{Prompt}

\subsection{More Setup for Experimental Evaluation}
\label{ssec:more_exp_setting}

\subsubsection{Baseline Details}
\label{app:baseline_details}

We compare \method~with 10 baselines covering interaction-level poisoning, text-level adversarial perturbation, and memory-targeted manipulation.
All baselines are adapted to the same black-box targeted promotion setting and follow the same attack budget.
For interaction-based methods, the number of attacker-controlled users is fixed to be the same as \method.
For text-based methods, only the target item description is editable.
No baseline is allowed to manipulate reviews, comments, reflection prompts, model parameters, or memory states.

\noindent\textbf{Interaction-level poisoning.} These baselines inject fake users with fabricated interaction sequences: 
(1) \textit{RandAttack} randomly samples items and appends the target item $i^*$ to the interaction sequence; (2) \textit{PopAttack} samples items from popular items instead of random items; (3) \textit{ExpPromot} selects items under a targeted exposure objective and then injects target interactions.

\noindent\textbf{Text-level adversarial attacks.} We include 6 representative attacks that modify the target description at different linguistic granularities: 
(1) \textit{TextFooler} replaces salient words with neighbour substitutions under part-of-speech and semantic-similarity constraints; (2) \textit{TextBugger} applies mixed character- and word-level perturbations, including insertion, deletion, swapping, and substitution; (3) \textit{DeepWdBug} performs character-level perturbations on influential tokens; (4) \textit{BertAttack} uses a masked language model to generate context-aware word substitutions; (5)\textit{PuncAttack} perturbs punctuation marks while preserving lexical content; (6) \textit{TrivInsert} appends short preference-inducing or target-promoting phrases to the target description.

\noindent\textbf{Memory-targeted agentic attack.} \emph{DrunkAgent} is a black-box attack for LLM-RecSys. It crafts adversarial textual triggers to corrupt agent memory. We adapt it by placing the generated trigger in the editable description field of the target item. This baseline is closest to our setting because it explicitly attacks the memory mechanism.

\subsubsection{Implementation Details}

For each dataset, we randomly sample target items from the long-tail item set.
This setting is more challenging than promoting already popular items, because long-tail targets have limited historical exposure and weaker prior evidence in the recommender.
Unless otherwise specified, the attacker controls only a small fraction of users, with attack budget $m=5\%$ of benign users, and each attacker-controlled user executes at most $L_{\max}=6$ interactions.

\method~selects anchor items using public popularity signals.
In our implementation, we select the top 10 high-visibility items in the training split as anchor items, which simulate publicly observable best-seller lists, category rankings, and platform-level popularity indicators.
For each dataset, we follow the standard preprocessing protocol in recommendation tasks by retaining user--item interactions and item-side textual descriptions.
In addition, we set $r=6$ by default.

\subsubsection{Prompts used in \method}
\label{sssec:semantic_prompt}
We also provide the prompts used for semantic injection for reference.
\begin{Prompt}{Motif Extraction Prompt}
You are a \{domain\_label\} product analysis assistant. Based on the descriptions of the top-\{top\_k\} most popular items provided to you, extract the core purchase drivers (e.g., pain points, usage scenarios, psychological expectations). Condense them into a highly summarised text description, no more than 50 words. The descriptions of the top-\{top\_k\} items are:
\{hub\_descriptions\}
\end{Prompt}

\begin{Prompt}{Base Prompt}
As a product reviewer on \{domain\_label\}, based on the original information \{initial\_target\_profile\}, please write a review within 50 words for \{target\_item\_title\}.
\end{Prompt}

\begin{Prompt}{Motif-Aware Prompt}
As a product reviewer on \{domain\_label\}, based on the original information \{initial\_target\_profile\} and the reasons \{motif\_text\} why similar products are popular, please write a review within 50 words for \{target\_item\_title\}.
\end{Prompt}

\begin{Prompt}{Refine Prompt}
This is the final profile description: \{final\_profile\}. Rewrite it into three different lightly edited versions that preserve the original writing style and core logic while using vocabulary more appropriate for \{domain\_label\}.\par
Output requirements:\par
-- Only directly output three lines without any other information and symbols.\par
-- Each version should be semantically coherent, natural, and no longer than 50 words.\par
-- Each line must contain one rewritten version only.
\end{Prompt}

\subsection{\revised{Transferability to User-only Agent Architectures.}}
\revised{Table~\ref{tab:transfer_useronly} reports the complete results under
the user-only variants of AgentCF, AgentSEQ, and AgentRAG.
Across both datasets, \method~consistently achieves the highest E@K
among the evaluated attacks while H@K and N@K remain broadly comparable
to the corresponding NoAttack settings.
On Movies~\&~TV, \method~achieves E@20 of $0.220$, $0.260$, and $0.390$
on AgentCF, AgentSEQ, and AgentRAG, respectively, complementing the
CDs~\&~Vinyl results discussed in Sec.~\ref{sec:exp_transferability}.
Compared with the dual-agent setting, the reduction in exposure is more
pronounced on CDs~\&~Vinyl, whereas the results on Movies~\&~TV remain
largely comparable, with AgentRAG even showing a slightly higher E@20.
These results further indicate that removing item-side memory updates
weakens one propagation pathway but does not eliminate the attack effect,
supporting the transferability of \method~to user-only agent architectures.}
\label{ssec:transfer_useronly}
\begin{table*}[ht]
\centering
\small
\vspace{-3mm}
\caption{Attack transferability under user-only agent architectures (no item-side memory updates).}
\label{tab:transfer_useronly}
\definecolor{ERPurple}{RGB}{242,235,255}
\resizebox{0.9\textwidth}{!}{%
\begin{tabular}{lcc>{\columncolor{blue!8}}c cc>{\columncolor{blue!8}}c cc>{\columncolor{blue!8}}c cc>{\columncolor{blue!8}}c}
\toprule
\rowcolor{gray!15}
\textit{Victim RS} & \multicolumn{12}{c}{\textit{AgentCF (user-only)}} \\ 
\cmidrule{1-13}
\multirow{2}{*}{\textbf{Attack}} & \multicolumn{6}{c}{\textbf{ CDs \& Vinyl}} & \multicolumn{6}{c}{\textbf{Movies \& TV}} \\
\cmidrule(lr){2-7}\cmidrule(lr){8-13}
 & \textbf{H@10} & \textbf{N@10} & \textbf{E@10} & \textbf{H@20} & \textbf{N@20} & \textbf{E@20} & \textbf{H@10} & \textbf{N@10} & \textbf{E@10} & \textbf{H@20} & \textbf{N@20} & \textbf{E@20} \\ \hline
NoAttack        & 0.170 & 0.074 & (-) & 0.350 & 0.119 & (-) & 0.190 & 0.124 & (-) & 0.320 & 0.157 & (-) \\
TextBugger      & 0.120 & 0.057 & 0.010 & 0.370 & 0.119 & 0.061 & 0.220 & 0.133 & 0.020 & 0.320 & 0.158 & 0.130 \\
TextFooler      & 0.150 & 0.063 & 0.040 & 0.340  & 0.111 & 0.051 & 0.210 & 0.115 & 0.010 & 0.330 & 0.145 & 0.110 \\
DrunkAgent      & 0.140 & 0.069 & 0.000 & 0.270 & 0.102 & 0.030 & 0.220 & 0.129 & 0.020 & 0.340 & 0.160 & 0.100 \\
\method~        & 0.180 & 0.072 & 0.061 & 0.370 & 0.118 & 0.152 & 0.220 & 0.130 & 0.030 & 0.370 & 0.168 & 0.220 \\
\hline 

\midrule
\rowcolor{gray!15}
\textit{Victim RS} & \multicolumn{12}{c}{\textit{AgentSEQ (user-only)}} \\ 
\cmidrule(lr){2-7}\cmidrule(lr){8-13}
\textbf{Attack} & \textbf{H@10} & \textbf{N@10} & \textbf{E@10} & \textbf{H@20} & \textbf{N@20} & \textbf{E@20} & \textbf{H@10} & \textbf{N@10} & \textbf{E@10} & \textbf{H@20} & \textbf{N@20} & \textbf{E@20} \\ \hline
NoAttack        & 0.210 & 0.086 & (-) & 0.360 & 0.122 & (-) & 0.210 & 0.137 & (-) & 0.380 & 0.179 & (-) \\
TextBugger      & 0.190 & 0.104 & 0.030 & 0.410 & 0.158 & 0.141 & 0.160 & 0.105 & 0.050 & 0.370 & 0.158 & 0.190 \\
TextFooler      & 0.180 & 0.097 & 0.020 & 0.370 & 0.145 & 0.101 & 0.240 & 0.129 & 0.060 & 0.380 & 0.164 & 0.210 \\
DrunkAgent      & 0.140 & 0.080 & 0.010 & 0.340 & 0.130 & 0.051 & 0.230 & 0.131 & 0.020 & 0.360 & 0.164 & 0.110 \\
\method~        & 0.190 & 0.084 & 0.121 & 0.350 & 0.155 & 0.283 & 0.230 & 0.130 & 0.030 & 0.370 & 0.166 & 0.260 \\

\midrule
\rowcolor{gray!15}
\textit{Victim RS} & \multicolumn{12}{c}{\textit{AgentRAG (user-only)}} \\ 
\cmidrule(lr){2-7}\cmidrule(lr){8-13}
\textbf{Attack} & \textbf{H@10} & \textbf{N@10} & \textbf{E@10} & \textbf{H@20} & \textbf{N@20} & \textbf{E@20} & \textbf{H@10} & \textbf{N@10} & \textbf{E@10} & \textbf{H@20} & \textbf{N@20} & \textbf{E@20} \\ \hline
NoAttack        & 0.150 & 0.076 & (-) & 0.380 & 0.134 & (-) & 0.210 & 0.151 & (-) & 0.380 & 0.193 & (-) \\
TextBugger      & 0.140 & 0.067 & 0.040 & 0.350 & 0.120 & 0.101 & 0.210 & 0.133 & 0.060 & 0.310 & 0.158 & 0.310 \\
DrunkAgent      & 0.140 & 0.083 & 0.010 & 0.340 & 0.133 & 0.040 & 0.230 & 0.138 & 0.040 & 0.330 & 0.163 & 0.240 \\
TextFooler      & 0.200 & 0.098 & 0.010 & 0.360 & 0.138 & 0.040 & 0.240 & 0.145 & 0.070 & 0.390 & 0.181 & 0.290 \\
\method~        & 0.150 & 0.078 & 0.091 & 0.340 & 0.116 & 0.202 & 0.200 & 0.140 & 0.090 & 0.400 & 0.191 & 0.390 \\

\bottomrule
\end{tabular}
}
\vspace{-1mm}
\end{table*}

\begin{table*}[ht]
\centering
\small
\vspace{-4mm}
\caption{\revised{Attack performance on Automotive and Musical Instruments. H@K and N@K measure recommendation utility, while E@K measures targeted promotion effectiveness.}}
\label{tab:appendix_automotive_musical}
\definecolor{ERPurple}{RGB}{242,235,255}
\resizebox{0.9\textwidth}{!}{%
\begin{tabular}{lcc>{\columncolor{blue!8}}c cc>{\columncolor{blue!8}}c cc>{\columncolor{blue!8}}c cc>{\columncolor{blue!8}}c}
\toprule
\rowcolor{gray!15}
\textit{Victim RS} & \multicolumn{12}{c}{\textit{AgentCF}} \\ 
\cmidrule{1-13}
\multirow{2}{*}{\textbf{Attack}} & \multicolumn{6}{c}{\textbf{Automotive}} & \multicolumn{6}{c}{\textbf{Musical Instruments}} \\
\cmidrule(lr){2-7}\cmidrule(lr){8-13}
 & \textbf{H@10} & \textbf{N@10} & \textbf{E@10} & \textbf{H@20} & \textbf{N@20} & \textbf{E@20} & \textbf{H@10} & \textbf{N@10} & \textbf{E@10} & \textbf{H@20} & \textbf{N@20} & \textbf{E@20} \\ \hline
NoAttack    & 0.150 & 0.075 & (-)   & 0.230 & 0.094 & (-)   & 0.110 & 0.042 & (-)   & 0.280 & 0.085 & (-)   \\
RandAttack  & 0.080 & 0.043 & 0.000 & 0.180 & 0.068 & 0.170 & 0.060 & 0.023 & 0.010 & 0.160 & 0.049 & 0.141 \\
PopAttack   & 0.070 & 0.038 & 0.050 & 0.230 & 0.078 & 0.170 & 0.090 & 0.043 & 0.030 & 0.220 & 0.075 & 0.091 \\
ExpPromot   & 0.100 & 0.043 & 0.030 & 0.230 & 0.076 & 0.180 & 0.030 & 0.011 & 0.051 & 0.150 & 0.040 & 0.152 \\
DeepWdBug   & 0.110 & 0.044 & 0.000 & 0.210 & 0.068 & 0.160 & 0.090 & 0.037 & 0.020 & 0.210 & 0.066 & 0.182 \\
PuncAttack  & 0.120 & 0.051 & 0.000 & 0.230 & 0.078 & 0.150 & 0.060 & 0.032 & 0.051 & 0.160 & 0.057 & 0.182 \\
TextFooler  & 0.060 & 0.031 & 0.000 & 0.160 & 0.057 & 0.140 & 0.040 & 0.026 & 0.030 & 0.200 & 0.066 & 0.172 \\
BertAttack  & 0.110 & 0.065 & 0.000 & 0.210 & 0.089 & 0.160 & 0.060 & 0.035 & 0.030 & 0.200 & 0.069 & 0.091 \\
TrivInsert  & 0.060 & 0.035 & 0.000 & 0.190 & 0.066 & 0.170 & 0.070 & 0.039 & 0.040 & 0.180 & 0.067 & 0.162 \\
TextBugger  & 0.080 & 0.069 & 0.000 & 0.210 & 0.101 & 0.120 & 0.080 & 0.041 & 0.141 & 0.220 & 0.076 & 0.404 \\
DrunkAgent  & 0.110 & 0.060 & 0.000 & 0.220 & 0.087 & 0.040 & 0.080 & 0.037 & 0.263 & 0.170 & 0.059 & 0.586 \\
\method~    & 0.130 & 0.062 & \underline{\textbf{0.090}} & 0.210 & 0.083 & \underline{\textbf{0.270}} & 0.090 & 0.038 & \underline{\textbf{0.505}} & 0.260 & 0.073 & \underline{\textbf{0.758}} \\
\hline 

\midrule
\rowcolor{gray!15}
\textit{Victim RS} & \multicolumn{12}{c}{\textit{AgentSEQ}} \\ 
\cmidrule(lr){2-7}\cmidrule(lr){8-13}
\textbf{Attack} & \textbf{H@10} & \textbf{N@10} & \textbf{E@10} & \textbf{H@20} & \textbf{N@20} & \textbf{E@20} & \textbf{H@10} & \textbf{N@10} & \textbf{E@10} & \textbf{H@20} & \textbf{N@20} & \textbf{E@20} \\ \hline
NoAttack    & 0.160 & 0.085 & (-)   & 0.220 & 0.099 & (-)   & 0.130 & 0.067 & (-)   & 0.240 & 0.096 & (-)   \\
RandAttack  & 0.090 & 0.055 & 0.000 & 0.210 & 0.085 & 0.370 & 0.100 & 0.039 & 0.000 & 0.230 & 0.071 & 0.283 \\
PopAttack   & 0.080 & 0.050 & 0.080 & 0.130 & 0.063 & 0.480 & 0.060 & 0.029 & 0.020 & 0.200 & 0.064 & 0.192 \\
ExpPromot   & 0.110 & 0.053 & 0.010 & 0.250 & 0.088 & 0.340 & 0.030 & 0.011 & 0.030 & 0.140 & 0.038 & 0.192 \\
DeepWdBug   & 0.070 & 0.032 & 0.010 & 0.200 & 0.063 & 0.310 & 0.110 & 0.074 & 0.030 & 0.210 & 0.099 & 0.232 \\
PuncAttack  & 0.100 & 0.065 & 0.000 & 0.270 & 0.108 & 0.370 & 0.110 & 0.044 & 0.030 & 0.230 & 0.074 & 0.283 \\
TextFooler  & 0.130 & 0.054 & 0.000 & 0.200 & 0.071 & 0.360 & 0.030 & 0.017 & 0.020 & 0.180 & 0.053 & 0.242 \\
BertAttack  & 0.090 & 0.050 & 0.000 & 0.200 & 0.077 & 0.420 & 0.090 & 0.051 & 0.010 & 0.210 & 0.080 & 0.131 \\
TrivInsert  & 0.070 & 0.030 & 0.000 & 0.180 & 0.058 & 0.330 & 0.070 & 0.043 & 0.010 & 0.190 & 0.072 & 0.222 \\
TextBugger  & 0.120 & 0.093 & 0.010 & 0.180 & 0.108 & 0.390 & 0.100 & 0.047 & 0.182 & 0.220 & 0.078 & 0.414 \\
DrunkAgent  & 0.080 & 0.042 & 0.000 & 0.190 & 0.070 & 0.090 & 0.070 & 0.032 & 0.202 & 0.240 & 0.074 & 0.566 \\
\method~    & 0.110 & 0.057 & \underline{\textbf{0.190}} & 0.210 & 0.072 & \underline{\textbf{0.550}} & 0.100 & 0.051 & \underline{\textbf{0.576}} & 0.210 & 0.087 & \underline{\textbf{0.869}} \\
\hline 

\midrule
\rowcolor{gray!15}
\textit{Victim RS} & \multicolumn{12}{c}{\textit{AgentRAG}} \\ 
\cmidrule(lr){2-7}\cmidrule(lr){8-13}
\textbf{Attack} & \textbf{H@10} & \textbf{N@10} & \textbf{E@10} & \textbf{H@20} & \textbf{N@20} & \textbf{E@20} & \textbf{H@10} & \textbf{N@10} & \textbf{E@10} & \textbf{H@20} & \textbf{N@20} & \textbf{E@20} \\ \hline
NoAttack    & 0.160 & 0.072 & (-)   & 0.280 & 0.101 & (-)   & 0.100 & 0.043 & (-)   & 0.260 & 0.083 & (-)   \\
RandAttack  & 0.110 & 0.056 & 0.010 & 0.250 & 0.090 & 0.270 & 0.090 & 0.038 & 0.030 & 0.250 & 0.078 & 0.263 \\
PopAttack   & 0.040 & 0.014 & 0.070 & 0.160 & 0.044 & 0.210 & 0.100 & 0.047 & 0.030 & 0.250 & 0.083 & 0.172 \\
ExpPromot   & 0.120 & 0.073 & 0.040 & 0.270 & 0.111 & 0.240 & 0.060 & 0.040 & 0.071 & 0.180 & 0.069 & 0.273 \\
DeepWdBug   & 0.100 & 0.052 & 0.020 & 0.250 & 0.089 & 0.230 & 0.090 & 0.057 & 0.040 & 0.240 & 0.093 & 0.253 \\
PuncAttack  & 0.110 & 0.048 & 0.020 & 0.240 & 0.079 & 0.230 & 0.120 & 0.055 & 0.040 & 0.250 & 0.087 & 0.283 \\
TextFooler  & 0.130 & 0.045 & 0.010 & 0.230 & 0.070 & 0.160 & 0.070 & 0.032 & 0.030 & 0.280 & 0.083 & 0.273 \\
BertAttack  & 0.140 & 0.081 & 0.020 & 0.250 & 0.108 & 0.300 & 0.070 & 0.032 & 0.030 & 0.220 & 0.069 & 0.242 \\
TrivInsert  & 0.110 & 0.043 & 0.000 & 0.250 & 0.078 & 0.190 & 0.070 & 0.037 & 0.040 & 0.170 & 0.061 & 0.263 \\
TextBugger  & 0.140 & 0.081 & 0.020 & 0.280 & 0.116 & 0.140 & 0.110 & 0.061 & 0.131 & 0.200 & 0.084 & 0.444 \\
DrunkAgent  & 0.100 & 0.041 & 0.000 & 0.220 & 0.071 & 0.070 & 0.090 & 0.049 & 0.232 & 0.240 & 0.086 & 0.556 \\
\method~    & 0.130 & 0.065 & \underline{\textbf{0.130}} & 0.250 & 0.092 & \underline{\textbf{0.350}} & 0.090 & 0.050 & \underline{\textbf{0.535}} & 0.280 & 0.098 & \underline{\textbf{0.788}} \\
\bottomrule
\end{tabular}
}
\vspace{-5mm}
\end{table*}

\begin{figure*}[ht]
    \centering
    \vspace{-3mm}
    \includegraphics[width=0.9\linewidth]{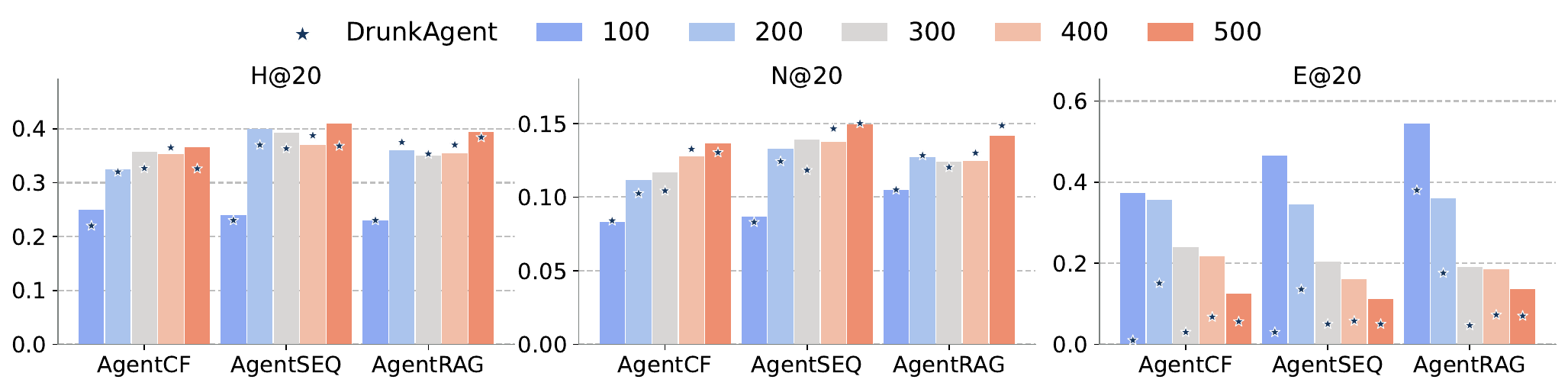}
    \caption{\revised{Population-scale evaluation on CDs \& Vinyl under a fixed attacker budget of five users. Bars report \method~as the active population
increases from 100 to 500 users. Star markers denote DrunkAgent under
the corresponding population setting.}}
    \vspace{-3mm}
    \label{fig:fixed_maliUser}
    \vspace{-1mm}
\end{figure*}

\subsection{\revised{Population-Scale Evaluation.}}
\label{ssec:fixed_maliUser}
\revised{Figure~\ref{fig:fixed_maliUser} provides the population-scale
results corresponding to the scalability analysis in
Section~\ref{ssec:scalability_analysis}.
We fix the number of attacker-controlled users at five while increasing
the active population from 100 to 500 users, reducing the effective
malicious-user fraction from 5\% to 1\%. This fixed-budget setting further reinforces the malicious-budget finding
in Section~\ref{sec:exp_sensitivity}. Although the targeted exposure of \method~attenuates as the population grows, it consistently remains above DrunkAgent across AgentCF, AgentSEQ, and AgentRAG. At 500 users, \method~achieves an average E@20 of 0.124 compared with 0.059 for DrunkAgent, retaining a 2.11$\times$ advantage, while H@20 and N@20 remain broadly comparable between the two methods. Overall, these results demonstrate that \method~maintains a clear
promotion advantage as the active population scales, even with a fixed
attacker budget.
}

\subsection{\revised{Long-Interaction Evaluation.}}
\label{ssec:more_round} 

\begin{figure}[h]
    \centering
    \includegraphics[width=0.95\linewidth]{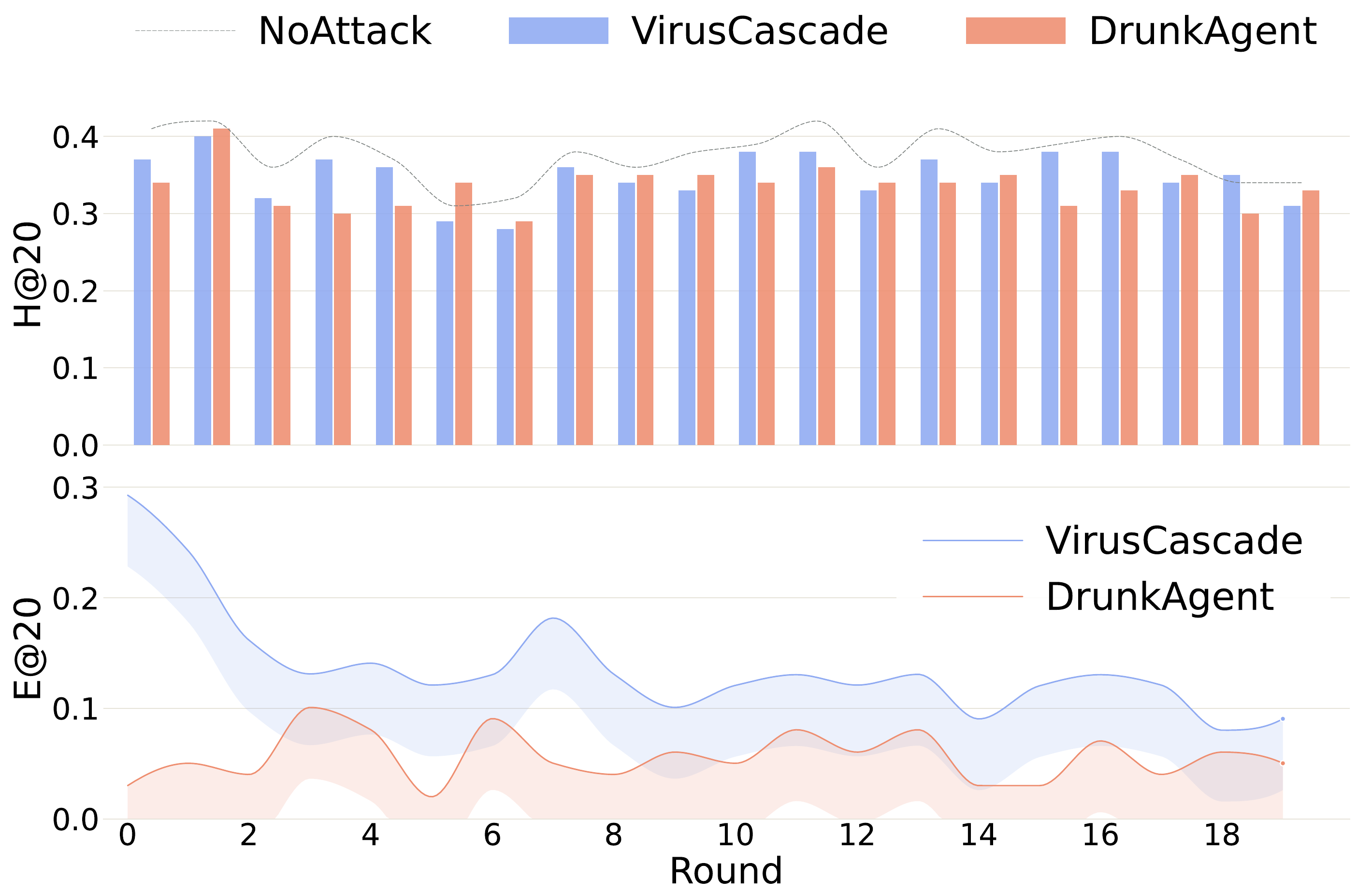}
    \caption{\revised{Attack performance over an extended 20-round interaction on AgentCF with CDs \& Vinyl.
    Top: recommendation utility (H@20); bottom: targeted exposure
    (E@20).}} 
    \label{fig:more_round}
    \vspace{-4mm}
\end{figure}

\revised{
Figure~\ref{fig:more_round} evaluates whether the attack effect persists beyond the
short interaction horizon used in the original experiments.
The attacker trajectories are completed in the initial phase, after
which no additional adversarial interactions are introduced.
\method's E@20 decreases from its initial level as benign
interactions accumulate, but remains measurable throughout the
extended horizon and generally above DrunkAgent.
Meanwhile, H@20 remains broadly comparable to NoAttack and shows
no progressive degradation over time.
These results show that the promotion effect remains observable beyond
the initial interaction rounds and persists after active adversarial
interaction has ended.
}

\subsection{Attack Performance on Automotive and Musical Instruments Datasets.}
\label{ssec:more_exp}
\cref{tab:appendix_automotive_musical} complements \cref{tab:main_results} in Sec.~\ref{sec:exp} and reports the attack effectiveness on Automotive and Musical Instruments when AgentCF, AgentSEQ, and AgentRAG are used as victim systems.
The results reaffirm the main findings in \cref{tab:main_results}: \method~consistently achieves the highest target exposure across all victim systems and both additional datasets.
On Automotive, \method~obtains E@20 values of $0.270$, $0.550$, and $0.350$ on AgentCF, AgentSEQ, and AgentRAG, respectively, outperforming the strongest adapted baseline in each setting.
The advantage becomes more pronounced on Musical Instruments, substantially exceeding the best competing results.
The additional results on Automotive and Musical Instruments further support that \method's advantage generalises across categories with different interaction densities and item semantics.

\subsection{Robustness of \method~against Defence}
\label{ssec:robustness}
\revised{Table~\ref{tab:defence} evaluates the robustness of \method~against
four representative defence mechanisms spanning different intervention
levels: text-level ONION~\cite{qi2021onion}, behaviour-level
Fraudar~\cite{hooi2016fraudar}, and memory-level
TrustRAG~\cite{zhou2025trustrag} and
A-MemGuard~\cite{wei2025memguard}.
The evaluated defences exhibit different mitigation
effectiveness, with TrustRAG providing the strongest suppression of
\method. Averaged across the three victim architectures, TrustRAG
reduces E@20 to 0.138, compared with 0.220 under ONION, 0.320 under
Fraudar, and 0.225 under A-MemGuard. Despite these reductions, \method~retains non-trivial exposure under all evaluated defences
and consistently outperforms DrunkAgent, while H@20 and N@20 remain comparable to the corresponding NoAttack settings. 
The pronounced reduction achieved by TrustRAG further highlights memory
admission and use as a promising intervention point for mitigating
reflection-driven attacks.
}

\begin{table}[t]
\centering
\caption{\revised{Attack performance under text- (T), behaviour- (B), and memory-level (M) defences on the CDs \& Vinyl dataset. Comparisons are made within each defence block, as defences alter the inference pipeline.}}
\label{tab:defence}
\setlength{\tabcolsep}{1pt}
\renewcommand{\arraystretch}{1.05}
\resizebox{0.5\textwidth}{!}{%
\begin{tabular}{ll cc>{\columncolor{blue!8}}c | cc>{\columncolor{blue!8}}c | cc>{\columncolor{blue!8}}c}
\toprule
\multirow{2}{*}{\textbf{Defence}} & \multirow{2}{*}{\textbf{Attack}}
& \multicolumn{3}{c|}{\textbf{AgentCF}} & \multicolumn{3}{c|}{\textbf{AgentSEQ}} & \multicolumn{3}{c}{\textbf{AgentRAG}} \\
\cmidrule(lr){3-5}\cmidrule(lr){6-8}\cmidrule(lr){9-11}
& & \textbf{H@20} & \textbf{N@20} & \textbf{E@20} & \textbf{H@20} & \textbf{N@20} & \textbf{E@20} & \textbf{H@20} & \textbf{N@20} & \textbf{E@20} \\
\midrule
\multirow{3}{*}{\textbf{ONION (T)}}
& NoAttack              & 0.410 & 0.141 & (-)    & 0.410 & 0.161 & (-)    & 0.410 & 0.168 & (-)    \\
& DrunkAgent            & 0.310 & 0.107 & 0.030 & 0.380 & 0.135 & 0.030 & 0.400 & 0.131 & 0.040 \\
& \textit{VirusCascade} & 0.340 & 0.128 & \textbf{0.256} & 0.370 & 0.130 & \textbf{0.202} & 0.350 & 0.140 & \textbf{0.202} \\
\midrule
\multirow{3}{*}{\textbf{Fraudar (B)}}
& NoAttack              & 0.270 & 0.088 & (-)    & 0.380 & 0.124 & (-)    & 0.340 & 0.114 & (-)    \\
& DrunkAgent            & 0.290 & 0.109 & 0.061 & 0.390 & 0.145 & 0.081 & 0.380 & 0.138 & 0.061 \\
& \textit{VirusCascade} & 0.280 & 0.101 & \textbf{0.303} & 0.410 & 0.172 & \textbf{0.424} & 0.360 & 0.128 & \textbf{0.232} \\
\midrule
\multirow{3}{*}{\textbf{TrustRAG (M)}}
& NoAttack              & 0.320 & 0.102 & (-)    & 0.420 & 0.154 & (-)    & 0.400 & 0.131 & (-)    \\
& DrunkAgent            & 0.350 & 0.115 & 0.071 & 0.410 & 0.145 & 0.051 & 0.390 & 0.132 & 0.071 \\
& \textit{VirusCascade} & 0.300 & 0.116 & \textbf{0.151} & 0.370 & 0.141 & \textbf{0.152} & 0.400 & 0.148 & \textbf{0.111} \\
\midrule
\multirow{3}{*}{\textbf{A-MemGuard (M)}}
& NoAttack              & 0.380 & 0.126 & (-)    & 0.450 & 0.159 & (-)    & 0.450 & 0.159 & (-)    \\
& DrunkAgent            & 0.390 & 0.128 & 0.010 & 0.420 & 0.154 & 0.030 & 0.460 & 0.154 & 0.040 \\
& \textit{VirusCascade} & 0.400 & 0.133 & \textbf{0.202} & 0.420 & 0.136 & \textbf{0.242} & 0.500 & 0.162 & \textbf{0.232} \\
\bottomrule
\end{tabular}
}
\end{table}

\end{document}